\documentclass[sigconf, nonacm]{acmart}
\usepackage{pvldb}
\usepackage{algorithmic}
\usepackage{algorithm}
\usepackage{graphicx}
\usepackage{textcomp}
\usepackage{xcolor}
\usepackage[linesnumbered,ruled,vlined,algo2e]{algorithm2e}
\usepackage{mathtools}
\usepackage{balance}
\usepackage{listings}
\usepackage{graphicx}
\usepackage{enumitem}
\usepackage{caption}
\usepackage{subcaption}
\usepackage{multirow}
\usepackage[linesnumbered,ruled,vlined,algo2e]{algorithm2e}
 
 \SetCommentSty{mycommfont} 
\usepackage{tikzscale}
\usepackage{tikz}
\usetikzlibrary{calc}
\usepackage{array}
\usetikzlibrary{shapes}
\usetikzlibrary{positioning}
\usetikzlibrary{arrows.meta}
\usepackage{circledsteps}
\usepackage{graphicx}
\usepackage{mathtools}
\usepackage{svg}
\usepackage[colorinlistoftodos,textsize=small]{todonotes}
\usepackage{xcolor}
\usepackage{xcolor}
\usepackage{amsmath}
\definecolor{siamgreen}{HTML}{159a88}
\tikzset{
blueNode/.style={circle,draw=black!50,fill=blue!20,thick,
inner sep=0pt,minimum size=3mm},
orangeNode/.style={circle,draw=black!50,fill=orange!80,thick,
inner sep=0pt,minimum size=3mm},
redNode/.style={circle,draw=black!50,fill=red!80,thick,
inner sep=0pt,minimum size=3mm},
}

\definecolor{deepgreen}{rgb}{0,0.7,0}

\providecommand{\coloredtcp}[1]{\tcp*{#1}}   
\providecommand{\coloredtcc}[1]{\tcp{#1}}    

\newcommand{\beamwidth}{beam width \ensuremath{\ell}\xspace}
\newcommand{\degreebound}{degree bound $\mathcal{R}$\xspace}

\newcommand{\ours}{{\texttt{GrAND}}\xspace}
\newcommand{\ourvamana}{{\ours-Vamana}\xspace}
\newcommand{\ourcagra}{{\ours-CAGRA}\xspace}
\newcommand{\ourfda}{{FreshDiskANN-GPU}\xspace}
\SetKwFor{ForPar}{foreach}{in parallel do}{}

\renewcommand\vldbdoi{XX.XX/XXX.XX}
\renewcommand\vldbpages{XXX-XXX}

\begin{document}
\title{\ours: GPU-based Dynamic Graph Indexes for Approximate Nearest Neighbour Search}

\author{Karthik Venkatasubba}
\orcid{0009-0001-6085-891X}
\affiliation{%
  \institution{IIT Hyderabad}
   \country{India}
}
\email{cs21resch14001@iith.ac.in}

\author{Shivendra Deshpande}
\affiliation{%
  \institution{IIT Hyderabad}
   \country{India}
}
\email{cs24mtech12017@iith.ac.in}

\author{Shivram S.}
\affiliation{%
  \institution{IIT Hyderabad}
   \country{India}
}
\email{ai24btech11031@iith.ac.in}

\author{Jyothi Vedurada}
\affiliation{%
  \institution{IIT Hyderabad}
   \country{India}
}
\email{jyothiv@cse.iith.ac.in}


\begin{abstract}
Modern Approximate Nearest Neighbour Search (ANNS) applications operate over continuously evolving vector collections and require graph indexes that sustain high-throughput searches while incorporating insertions and deletions with high recall. However, most GPU graph indexes are static or provide limited update support. Updates require neighbour discovery, reverse-edge creation, pruning, and deletion-induced graph repair; executing these operations concurrently introduces redundant distance computations and conflicting accesses to shared adjacency lists. Background-rebuild-based deletion further incurs substantial computation, additional memory consumption, and interference with foreground queries.

We present \ours (GPU-based Dynamic Graph Indexes for Approximate Nearest Neighbour Search), a GPU-native collection of dynamic-update algorithms for two popular graph indexes, Vamana and CAGRA. \ours consolidates graph repair across a batch, eliminating redundant pruning computations, and employs a lock-free find-and-replace strategy for parallel adjacency-list updates. For reliable in-place deletion, \ours constructs an on-demand reverse graph on the GPU, accurately identifying incoming edges without permanently duplicating the index. We evaluate \ours on seven real-world datasets across five streaming workloads, comparing it against SVFusion and FreshDiskANN-GPU (our GPU adaptation of FreshDiskANN). 
\ours improves overall workload throughput by 2.2$\times$--8.7$\times$ and 6.5$\times$--25.4$\times$, respectively, while maintaining high search throughput and recall over sustained updates.

\end{abstract}
\maketitle

\vldbtopmatter


\section{Introduction}
\label{sec:introduction}

Modern data-intensive applications represent objects such as text,
images, videos, user preferences, and sensor observations as
high-dimensional vector embeddings~\cite{lu2025multimodal}. 
Given a query vector, $k$-nearest-neighbour search identifies the $k$ 
 most similar dataset vectors under a chosen distance metric. 
Since exhaustive search requires comparing the query against every dataset vector, it becomes prohibitively expensive for large, high-dimensional datasets, motivating approximate nearest-neighbour search (ANNS), which trades a small loss in recall for substantially lower search cost~\cite{indyk1998ann,wang2021graphsurvey}.
Consequently, ANNS indexes have become the dominant choice for vector databases and
underpin applications such as semantic search, retrieval-augmented generation,
recommendation systems, computer vision, and multimedia
retrieval~\cite{milvus2021,lewis2020rag,covington2016youtube}.

\begin{figure}[t]
    \centering
    \includegraphics[width=0.85\columnwidth, trim={3.5cm, 0cm, 3.5cm, 0cm}]{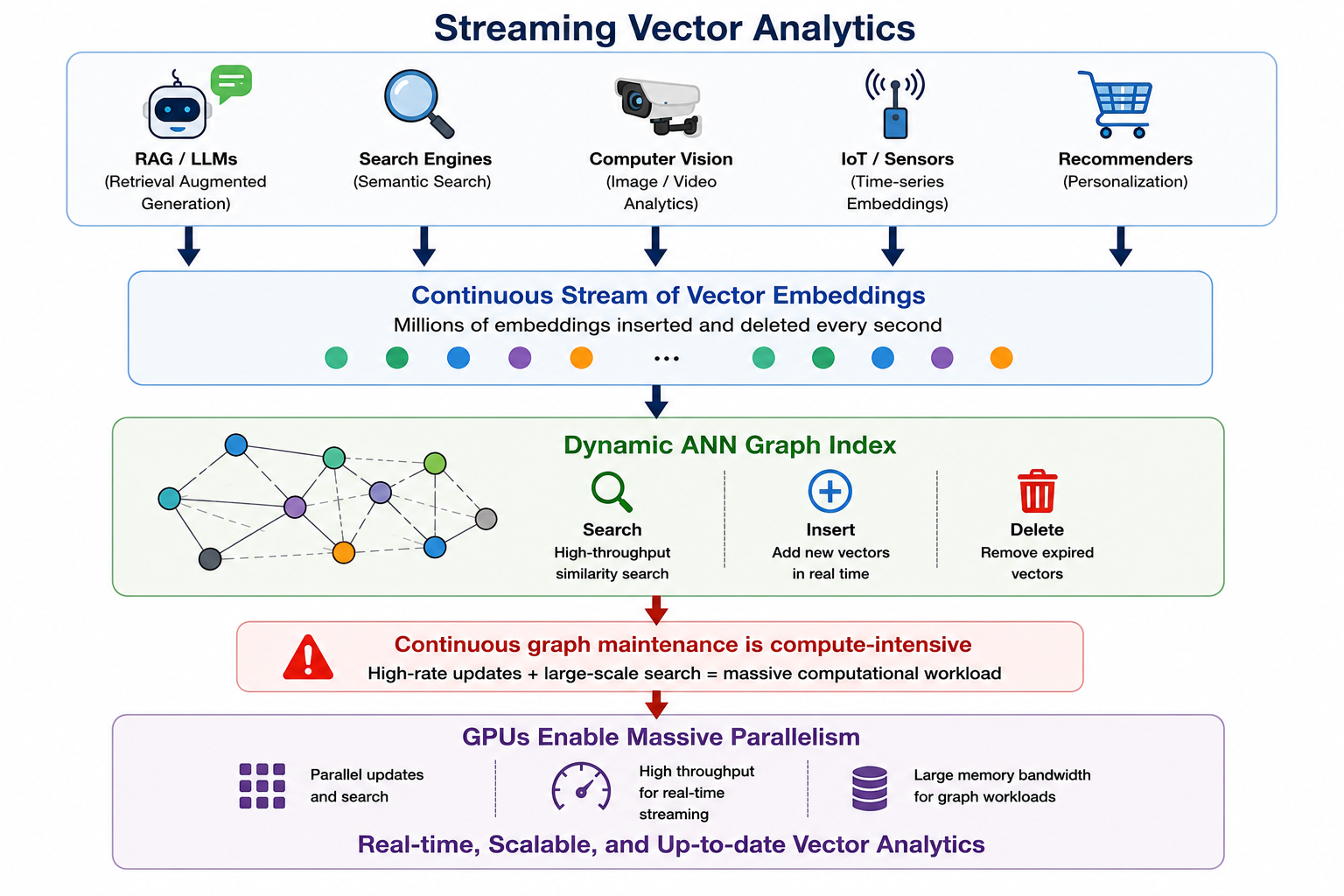}
    \caption{Modern AI applications generate streams of vectors that
    require dynamically maintained ANN graph indexes.}
    \label{fig:streaming_block_diagram}
\end{figure}

Most high-performance ANNS indexes~\cite{BANG, ootomo2024cagra}, however, are designed primarily for static datasets. 
This assumption is increasingly unsuitable for
applications in which new embeddings arrive continuously, existing objects
are updated, and obsolete data must be removed. As illustrated in
Figure~\ref{fig:streaming_block_diagram}, a dynamic ANNS index must
interleave high-throughput searches with concurrent insertions and deletions while
preserving index freshness and search recall~\cite{freshdiskann}. 
Consequently, there has been continued interest in exploring hardware acceleration, e.g. GPUs, for vector databases~\cite{stonebraker2024goes}.
However, supporting dynamic indexes on GPUs introduces several key challenges, in particular for graph-based indexes, the dominant ANNS index type, which requires complex graph maintenance for every update.


\noindent\textit{Challenge 1. Efficient Graph Repair.}
In dynamic graph processing, each insertion or deletion affects multiple neighbourhoods and requires neighbour discovery, reverse-edge updates, pruning to enforce the degree bound, and repair of adjacency lists affected by deletions in order to preserve graph quality.
These operations are computationally expensive because they involve numerous distance computations over candidate vectors. 
Moreover, pruning may be performed during both forward-edge construction and subsequently during reverse-edge updates,  leading to redundant work when the candidate sets overlap.
Performing these operations in a massively parallel manner on GPUs also introduces race conditions and complex synchronisation requirements because multiple threads may concurrently access or modify the same adjacency lists.

Furthermore, most existing techniques~\cite{peng_svfusion_2026,freshdiskann} handle deletions via tombstoning, deferring the actual deletion to a background index rebuild/consolidation operation, which removes the deleted node from the graph index and repairs the adjacency lists of the affected nodes eventually. 
This approach comes with undesirable side effects:\,(i) since rebuild is a computationally intensive task, during this background activity, the foreground query processing is starved, resulting in lower throughputs as evidenced by our experiments;
(ii) index rebuild requires additional temporary memory comparable to the size of the current index, increasing the peak memory requirement and limiting scalability to larger datasets (e.g., 100M vectors) on memory-constrained GPUs.


\noindent\textit{Challenge 2. Efficient In-place Deletion.}
An alternative to existing dynamic ANNS systems that handle deletions via background index rebuilding~\cite{freshdiskann} is to support in-place deletions.
However, graph-based ANNS indexes maintain only the outgoing adjacency list of each node. 
Consequently, when a node is deleted, the index lacks sufficient information to instantly identify all nodes whose adjacency lists contain an edge to the deleted node (i.e., the deleted node's incoming edges). 
Therefore, existing dynamic ANNS systems support in-place deletions~\cite{xu2025place} using local-neighbourhood heuristics that approximate the set of incoming edges. 
However, these approaches incur substantial computational/memory overhead and fail to identify every incoming edge that subsequently affects recall over prolonged operations.

Existing GPU-based graph ANNS systems do not fully address the challenges discussed above.
CAGRA~\cite{ootomo2024cagra} and BANG~\cite{BANG} target static GPU indexes, Jasper~\cite{jasper}  supports only batch-incremental insertions but not streaming deletions, and SVFusion~\cite{peng_svfusion_2026} relies on tombstone-based deletions without in-place graph maintenance and further requires CPU-GPU collaboration overheads.
Thus, efficient support for continuous insertions and deletions within a GPU-resident graph
index while preserving graph quality and high GPU throughput remains unaddressed.

To address these challenges and limitations, we present \ours, an ANNS system for
streaming vectors built on four key innovations.
(i) Full GPU-native: \ours is optimised for dynamic update processing to maximise throughput for datasets that fit in GPU memory, without relying on CPU-side graph processing.
(ii) Efficient Index Repair: instead of separately
pruning affected nodes during each insertion or
deletion, \ours{} accumulates the affected nodes and prunes them collectively
at the end of the batch, eliminating redundant computations
without compromising recall. 
To achieve maximum parallelism during pruning, which involves concurrent reads and writes of adjacency lists, we employ a simple
find-and-replace strategy to update nodes in the adjacency list, thereby avoiding fine-grained locking.
(iii) Accurate and Efficient In-place Deletions: unlike heuristic approaches that approximate incoming/reverse-edges, 
\ours exploits GPU parallelism and memory bandwidth to construct a reverse graph for each batch of deletions, enabling accurate identification of incoming edges and in-place repair of affected adjacency lists without maintaining a permanent reverse graph.
(iv) Extensible ANNS Algorithms: the dynamic-update techniques implemented by \ours can be applied to both Vamana~\cite{diskann} and CAGRA~\cite{ootomo2024cagra}, two popular graph-based indexes. 

Together, these techniques deliver an overall throughput improvement of 2.2$\times$--8.7$\times$ over the state-of-the-art SVFusion baseline while maintaining high recall.

The main contributions of this paper are as follows:
\begin{itemize}
    \item We present \ours, a collection of high-performance dynamic update
    algorithms that extend Vamana and CAGRA graph indexes with efficient support for streaming
    insertions and deletions.


    \item We design a robust, lock-free graph-repair strategy that
    consolidates pruning across affected vertices within an update batch,
    eliminating redundant distance computations while enabling
    massively parallel GPU execution.

    \item We propose a novel GPU-friendly on-demand reverse graph construction
    technique that accurately identifies incoming edges, enabling
    reliable and efficient in-place deletion.

    \item We conduct an extensive evaluation using seven real-world
    datasets under five diverse workload patterns. 
    Compared with SVFusion and FreshDiskANN-GPU (our GPU adaptation of FreshDiskANN~\cite{freshdiskann}), \ours\ achieves 2.5$\times$--20.2$\times$ and 5.0$\times$--39.8$\times$ higher insertion throughput, respectively, and 2.2$\times$--8.7$\times$ and 6.5$\times$--25.4$\times$ higher overall throughput, while maintaining comparable search throughput and recall sustained throughout update operations.
\end{itemize}

\section{Background}
\label{s:back}

In this section, we introduce the core primitives of graph-based ANNS indices, discuss two popular indexing techniques, Vamana~\cite{diskann} and CAGRA~\cite{ootomo2024cagra}, which underpin our dynamic GPU index construction, and summarise the GPU concepts needed for parallelisation.


\begin{algorithm2e}[t]
  \small
  \DontPrintSemicolon
  \SetNoFillComment

  \caption{BeamSearch$(G,s,q,\ell,k, \mathrm{Dist})$}
  \label{alg:beam-search}

  \KwIn{Proximity graph $G$; starting point $s$; query vector $q$; \beamwidth; recall parameter $k$; distance function $\mathrm{Dist}$ }
  \KwOut{The $k$ closest vertices to $q$ visited during the search}

  $V \gets \emptyset$;\quad
  $C \gets \{s\}$\;

  \While{$C \setminus V \ne \emptyset$}{
    $u \gets
      \arg\min_{v \in C \setminus V}  \mathrm{Dist}(v,q)$\;

    $V \gets V \cup \{u\}$\;

    $C \gets C\ \cup N_{out}(u)$\;

    \If{$|C| > \ell$}{
      $C \gets \text{the $\ell$ vertices in $C$ closest to $q$}$\;
    }
  }

\Return{the $k$ vertices in $\mathcal{C}$ closest to $q$}\;
\end{algorithm2e}

\subsection{Graph-based Index Primitives}
Typically, a proximity graph index is constructed from the dataset using a distance metric (e.g., Euclidean distance) between vectors~\cite{liu2026gpu}. 
The index is a directed graph where each node stores its outgoing neighbours and is subject to a degree bound.

\noindent\textbf{Beam Search:}
Graph-based ANNS performs approximate nearest neighbour search through BeamSearch (see Algorithm~\ref{alg:beam-search}), which traverses the graph to identify vectors closest to a query.
Starting from a designated node $s$, we fetch its neighbours and maintain a sorted worklist $C$ of size $\ell$ based on their distances to the query $q$. A visited set $V$ prevents repeated exploration and duplicate entries in the worklist. 
Each iteration expands the closest unvisited candidate until no closer candidates are found. Finally, the first $k$ entries in the worklist are returned as the query's nearest neighbours.

\noindent\textbf{Pruning:}
During index construction, a new node's adjacency list is built using beam search to identify nearby nodes and establish forward and backward edges.
This may produce more candidates than the degree bound permits. 
The Prune Algorithm~\ref{alg:prune} selects the best-suited $\mathcal{R}$ candidates from the candidate set $C$.
It iteratively scans the candidate set $C$, identifies the closest candidate $p^\ast$ using the index-specific $\textit{}{Nearest}()$ function, and adds it to the adjacency list $L$.
This newly established edge may make some candidates in $C$ redundant, and can be identified using an index-specific $\texttt{}{Redundant}$ function based on pairwise distance computations.
The process terminates when $L$ contains $\mathcal{R}$ entries or $C$ is empty.



\subsection{Two Popular Graph-based ANN Indexes}
\noindent\textbf{Vamana:} 
The Vamana index~\cite{diskann} implements the BeamSearch and Prune primitives through the \textit{GreedySearch} and \textit{RobustPrune} routines, respectively, as described in DiskANN~\cite{diskann}. 
Both routines are computationally intensive due to their repeated distance computations. 
\textit{RobustPrune} removes redundant local edges while preserving diverse long-range connections, producing a sparse (with variable node degrees) yet highly navigable graph with the small-world property.
Consequently, ANN search greedily traverses the graph from a designated entry point, typically the graph medoid, to efficiently converge on the query's nearest neighbours with high search accuracy.
BANG~\cite{BANG} presents an optimised GPU implementation of \textit{GreedySearch}, which we refer to as \textit{BANGSearch}.


\noindent\textbf{CAGRA:} CAGRA~\cite{ootomo2024cagra} is a GPU-oriented graph-based ANNS index that represents each data point as a vertex with a fixed number of outgoing edges. 
Its construction first generates an approximate k-NN graph~\cite{wang2021fast}, typically using the highly parallel NN-Descent~\cite{dong2011efficient} algorithm, and subsequently optimises it through rank-based edge reordering, pruning and reverse-edge addition. These transformations remove redundant local paths, improve graph connectivity and increase the number of vertices reachable within a small number of hops. 
The resulting fixed-degree, directed proximity graph provides regular computation and exposes substantial parallelism, making it well-suited for graph traversal on GPUs. 
We refer to the optimised GPU implementation of its search primitive~\cite{ootomo2024cagra} as \textit{CAGRASearch}.

\subsection{GPU Programming Model}

\begin{algorithm2e}[t]
  \small
  \DontPrintSemicolon
  \SetNoFillComment
  \SetKw{Return}{return}
  \SetKw{Break}{break}

  \caption{Prune$(G,p,C,\mathcal{R})$}
  \label{alg:prune}

  \KwIn{Graph $G$; node $p$; candidate set $C$;
        \degreebound \\
       \quad \quad Index-specific Helper functions: $\mathrm{Nearest()}$ and 
        $\mathrm{Redundant()}$}

  \KwOut{Pruned adjacency list $\mathcal{L}$ for $p$}

  $\mathcal{L}\gets\emptyset$\;

  \While{$C\ne\emptyset$}{
    $p^\ast\gets\mathrm{Nearest}(G,p,C)$\label{ln:prune_nrearest}
      \tcp{Select the nearest node to $p$ from the current candidates }

    $\mathcal{L}\gets\mathcal{L}\cup\{p^\ast\}$

    \If{$|\mathcal{L}|=\mathcal{R}$}{
      \Break
    }

    $T\gets
      \mathrm{Redundant}(G,p,p^\ast,\mathcal{L},C)$ \label{ln:prune_redundant}
      \tcp{Find candidates that can be pruned based on their distances to $p$ and $p^\ast$}

    $C\gets C\setminus\bigl(T\cup\{p^\ast\}\bigr)$
      \tcp{Discard processed and redundant candidates}
  }

  \Return{$\mathcal{L}$}\;
\end{algorithm2e}

Our algorithms are implemented in CUDA~\cite{cuda,cudablog}, which organises computation into threads, thread blocks, and grids.
Threads in a thread block cooperate via shared memory and \texttt{\_\_syncthreads()}, while thread blocks execute independently. A \emph{warp} of 32 threads executes in lock-step (SIMT), so minimising branch divergence is essential for throughput. We coordinate concurrent updates to shared structures using \texttt{atomicAdd()} and \texttt{atomicCAS()} rather than locks, and overlap kernels and transfers across multiple \emph{CUDA streams} to improve resource utilisation.

\section{Overview and Preliminaries}
\label{s:overview}

This section presents an overview of the design of \ours\ and the preliminaries required to describe our algorithms.



\begin{figure*}[t]
    \centering
    \includegraphics[width=1\linewidth]{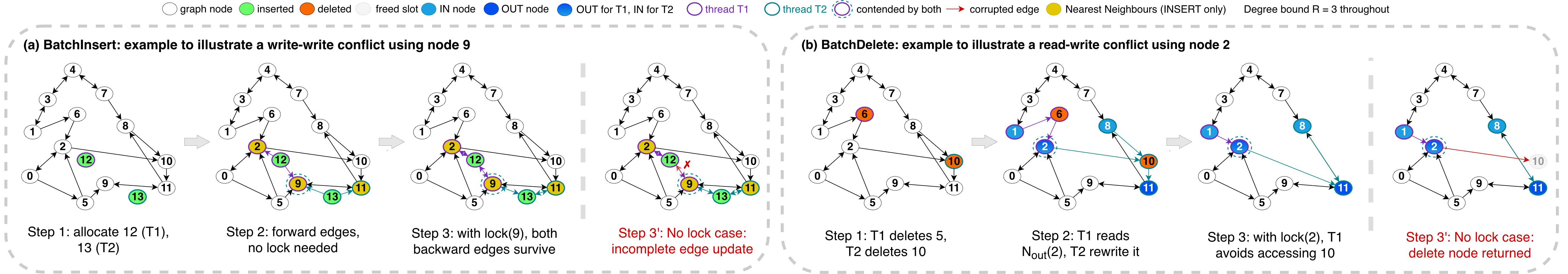}
    \caption{High-level steps in handling insert (a) and delete (b) operations on graph index via a toy example illustration.}
    \label{fig:toy_example}
\end{figure*}

\subsection{Overview}
\label{ss:challenges}
We identify the key challenges in supporting dynamic updates to graph indexes entirely on GPUs and propose efficient solutions that deliver high query performance, incur low update overhead, and preserve high recall.

\noindent\textbf{Robust and Lock-free Graph Index Repair/Pruning:}
Supporting dynamic updates in graph-based  ANNS indexes on GPUs is challenging because multiple insertions and deletions execute the pruning operation concurrently and frequently modify overlapping graph neighbourhoods. Figure~\ref{fig:toy_example} illustrates the two primary forms of conflicts encountered during the graph index updates. 
Figure~\ref{fig:toy_example}(a) shows a write-write conflict during concurrent insertions, where two threads simultaneously insert nodes 12 (T1) and 13 (T2). 
While forward-edge construction is conflict-free since each thread updates only its newly allocated node (Step~2), the conflict arises during reverse-edge installation when both threads simultaneously update the adjacency list of node~9. 
With synchronisation (Step~3), node~9 correctly retains all reverse edges to $\{11,12,13\}$. 
However, without synchronisation (Step~3'), during concurrent updates by T1 and T2 while modifying the adjacency list of node~9, T2's update overwrites T1's update (\emph{write-write race condition}), causing the edge $9 \rightarrow 12$ to be lost. 
Figure~\ref{fig:toy_example}(b) illustrates a read-write conflict during concurrent deletions, where thread T1 reads the adjacency list of node 2 while thread T2 simultaneously repairs it. 
Restricting to the time window of concurrent access of node~2 by T1 and T2, without synchronisation (Step~3'), the deleted edge $2 \rightarrow 10$ may be read by T1 while T2 is simultaneously updating/repairing the adjacency list of node~2. 
With synchronisation (Step~3), T2 has to wait for T1's access of node 2 before it can delete node 10 and vice versa. T1 reading a deleted node 10 via node 2's adjacency can potentially be avoided. 
Such write-write and read-write conflicts become increasingly common when thousands of GPU threads concurrently access overlapping adjacency lists. 
Although lock-based synchronisation using atomic operations can preserve correctness, it incurs significant contention and busy waiting, limiting scalability.
To effectively address this challenge, we develop a conflict-free node-level update strategy. 
Multiple adjacency list updates required for a single insert or delete operation are first localised, and we identify all affected nodes. Then we read the respective adjacency lists and stage the delta updates in local buffers for each adjacency list. Finally, for each affected node, we assign one thread to merge the delta updates and compute the resulting adjacency list. Since each thread is updating only one adjacency list, there are no conflicts.

 Pruning involves a significant amount of distance computations as part of the \texttt{Nearest} and \texttt{Redundant} functions. 
In the Vamana index, although these computations are unavoidable, we identify redundant distance computations between the same pair of points across Lines~\ref{ln:prune_nrearest} and \ref{ln:prune_redundant} in Algorithm~\ref{alg:prune}, and reuse the previously computed distances across iterations.
 Interestingly, for the CAGRA index, we perform pruning using a find-and-replace strategy (with the neighbour-of-neighbour as the neighbour property), avoiding distance computations altogether.


\noindent\textbf{Efficient and Accurate In-place Deletion:}
Processing node deletions in-place is preferred over lazy deletes, as the latter leads to gradual degradation in recall over time, necessitating periodic index rebuilds.
Deleting a node in-place from a graph-based ANN index requires identifying all nodes with edges pointing to the deleted node $p$ (incoming edges of $p$) so that their adjacency lists can be repaired to replace references to $p$, thereby preserving graph connectivity and search quality.
This is challenging because the adjacency list stores references to the outgoing edges (OUT nodes), not incoming edges (IN nodes), to keep the index size compact.
Consequently, identifying IN nodes of a given node to be deleted requires scanning and comparing all adjacency lists, which is prohibitively expensive. 
As noted in prior work~\cite{freshdiskann,xu2025place,zhang_cleanann_2026}, implementing an accurate and efficient deletion strategy is therefore inherently difficult.
Existing systems, therefore, rely on heuristics to approximate incoming edges (e.g., IP-DiskANN~\cite{xu2025place}). 
This approach involves intensive graph traversals, making it computationally heavy.
To address these challenges, we explore the idea of a reverse graph, i.e., a data structure that allows us to readily look up the accurate IN nodes of a node. 
We generate the reverse graph on demand with minimal latency by leveraging GPU parallelism and optimised memory access.  



\noindent\textbf{Efficient resource utilisation:}
Streaming data involves frequent insertions and deletions and therefore requires a strategy to efficiently utilise the limited memory resources~\cite{Sun2024}. 
The preliminary step in processing any streaming workload is allocating and deallocating memory. 
Conventionally, we would allocate/deallocate inline for each insert/delete operation in the input stream, which leads to memory fragmentation and performance overhead. During the insert flow, SVFusion allocates device memory for the dataset/graph expansion.  
Understandably, this is computationally costly, especially as the batch sizes increase. 
Frequent memory allocations and deallocations are not heap-friendly (i.e., they lead to fragmentation). 
So, we overcome this challenge through a resource-pool reuse strategy, where we allocate memory during initialisation to a preconfigured maximum peak value and reuse this pool effectively throughout the insert/delete lifecycle. This avoids frequent allocations and deallocations, which helps to improve the performance.

\subsection{Preliminaries}
\label{ss:overview}

This section presents high-level, generic algorithms to support insertions and deletions during dynamic index updates. 
These generic algorithms provide the foundation for the specific instantiations of \ours{} for Vamana and CAGRA graph indexes, 
\ourvamana and \ourcagra described in Section~\ref{sec:grand-vamana-cagra}. 
We omit the search algorithms because \ours\  directly inherits them from the respective implementations in ~\cite{BANG} and~\cite{ootomo2024cagra}.


\begin{algorithm2e}[t]
  \small \DontPrintSemicolon \SetNoFillComment
\caption{Generic BatchInsert and BatchDelete Operations}
  \label{alg:generic}
  \SetKwProg{Proc}{Procedure}{}{}
  \SetKwFunction{FBatchInsert}{BatchInsert}\SetKwFunction{FBatchDelete}{BatchDelete}\SetKwFunction{FSearch}{Search}
  \SetKwFunction{FPrune}{Prune}\SetKwFunction{FInNbrs}{InNeighbuors}


  \Proc{\FBatchInsert{$\mathcal{B}$, $G$,  $\mathcal{R}$}}{
    \KwIn{Batch $\mathcal{B} = \{(id_i, x_i)\}_{i=1}^{B}$; Graph $G$; degree bound $\mathcal{R}$}
  \KwOut{Graph $G$ updated to include all points in $\mathcal{B}$}
    \ForPar{$(id_i, x_i) \in \mathcal{B}$}{
      copy vector $x_i$ into dataset;
      allocate node $p_i$ for $x_i$
    }
    
    \ForPar{$p_i\in\mathcal{B}$}{
      $(L_i, V_i) \gets$ \FSearch{$G, x_i$} \coloredtcp{candidate set}
      $N_{\mathrm{out}}(p_i) \gets$ \FPrune{$p_i, V_i, \mathcal{R}$} \coloredtcp{forward edges}
    }

  \ForPar{$p_i\in\mathcal{B}$}{
        \ForEach{$u \in N_{\mathrm{out}}(p_i)$}{
            $N_{\mathrm{out}}(u) \gets N_{\mathrm{out}}(u) \cup \{p_i\}$ \coloredtcp{reverse edge}
            \If{$|N_{\mathrm{out}}(u)| > \mathcal{R}$}{
                $N_{\mathrm{out}}(u) \gets$ \FPrune{$u, N_{\mathrm{out}}(u), \mathcal{R}$}\;
            }
        }
    }
  }
  \Proc{\FBatchDelete{$\mathcal{B}$, $G$,  $\mathcal{R}$}}{
    \KwIn{deletions $\mathcal{B} = \{p_i\}_{i=1}^{B}$;  Graph $G$, \degreebound}
\KwOut{Graph $G$ updated after deleting all points in $\mathcal{B}$}
    \ForPar{$p_i \in \mathcal{B}$}{
      $I_i \gets$ $N_{in}(p_i)$ \coloredtcp{finding incoming edges}
    }

    \ForPar{$p_i \in \mathcal{B}$}{ \label{alg:generic-delete-start}
        \ForPar{$z \in I_i$}{
            adjust $N_{\mathrm{out}}(z)$ \coloredtcp{adjusting in-neighbours of $p_i$}
        }
    }

    \ForPar{$p_i \in \mathcal{B}$}{ 
        \ForPar{$z \in N_{\mathrm{out}}(p_i)$}{
            adjust $N_{\mathrm{in}}(z)$ \coloredtcp{adjusting out-neighbours of $p_i$} \label{alg:generic-delete-end}

        }
    }

    \ForPar{$p_i \in \mathcal{B}$}{
      remove $p_i$ from $G$\; 
    }
  }

\end{algorithm2e}

\noindent\textbf{Generic Insert:}
Algorithm~\ref{alg:generic} presents a generic procedure for parallel batch insertion into a graph-based ANN index. 
It captures the common sequence of operations performed by dynamic graph indexes while abstracting index-specific search, neighbourhood construction and synchronisation aspects. 

The insertion procedure then processes each vector in the batch independently in parallel. For each vector, the abstract \textit{Search} primitive computes a candidate neighbourhood around the query vector. 
Depending on the underlying index, \textit{Search} may be implemented using greedy graph traversal, beam search, or any other graph exploration strategy. 
The candidate set is then processed by the abstract \textit{Prune} routine to construct the outgoing adjacency list while enforcing the degree bound $\mathcal{R}$. 
Since each thread updates only the adjacency list of its own newly allocated node, forward-edge construction is conflict-free and requires no synchronisation.
Once all outgoing (forward) edges have been constructed, reverse edges are added by inserting each new node into the adjacency lists of its selected neighbours. Unlike forward-edge construction, this step updates existing nodes that may be shared across multiple insertions, introducing concurrent accesses that require synchronisation. 
The \textit{Prune} procedure is applied to every affected node, and it is possible that the same node is subject to pruning in parallel.


 Figure~\ref{fig:toy_example}(a) illustrates the high-level steps of insertion using a toy example. 
The graph has a degree bound $\mathcal{R}=3$, and nodes 12 and 13 are new vectors to be inserted by threads T1 and T2, respectively. As the first step, nodes 12 and 13 are initialised. 
Next, the algorithm establishes the outgoing (forward) edges for both nodes independently without synchronisation, since each thread updates only its own adjacency list. 
Finally, reverse edges are added by updating the adjacency lists of the selected neighbours, e.g. edge $9 \rightarrow 12$ in Step 3.

\noindent\textbf{Generic Delete:}
Algorithm~\ref{alg:generic} also presents a generic procedure for parallel batch deletion from a graph-based ANN index. 
Given a batch of nodes to be deleted, the procedure first identifies their incoming neighbours through the abstract \textit{InNeighbors} primitive. 
Depending on the index, incoming neighbours may be retrieved directly from an explicit reverse graph or discovered through repair traversals or other auxiliary mechanisms when only outgoing adjacency is maintained.
Once the incoming neighbourhoods have been identified, the procedure repairs the graph by rewiring the local neighbourhoods around each deleted node (Lines \ref{alg:generic-delete-start}-\ref{alg:generic-delete-end}). 
Specifically, for every incoming and outgoing neighbour of a deleted node, the corresponding adjacency list is updated by removing references to the deleted node and introducing connections to index-specific replacement nodes. 
As in batch insertion, these updates modify adjacency lists shared across multiple concurrent deletions and therefore require synchronisation. 
After rewiring, any neighbourhood whose degree exceeds the degree bound $\mathcal{R}$ may be processed by the abstract \textit{Prune} routine to restore the degree constraint. 

Figure~\ref{fig:toy_example}(b) illustrates the high-level steps in deletion using a toy example. 
Nodes 6 and 10 are to be deleted concurrently by threads T1 and T2, respectively. 
The algorithm first identifies the incoming (IN) and outgoing (OUT) neighbours/edge of each deleted node, then repairs the affected adjacency lists by removing references to the deleted nodes and adding replacement edges.
In this example, while T1 updates the adjacency list of node~2 during the deletion of node~6, T2 simultaneously reads the same adjacency list during the deletion of node~10, resulting in a conflicting access.

\section{G\lowercase{r}AND}
\label{s:tech}
\label{sec:grand-vamana-cagra}
Extending on the generic algorithms presented in the previous section, we now present  
\ourvamana and \ourcagra, the specific instantiations for the Vamana and CAGRA graph indexes.
During initialisation, GrAND allocates GPU memory to persist the base dataset and the graph index, accommodating a preconfigured peak size of $N$ elements. The corresponding vector or adjacency list can be looked up using an ID in the range $[0, N-1]$. Since we support in-place deletions, the deleted ID can be reused in future insertions. The batched operation is executed in parallel by multiple thread-blocks, with each thread-block processing one input node in parallel.  


\subsection{\ourvamana}
\label{sec:grand-vamana}

\begin{algorithm2e}[t]
\small \DontPrintSemicolon \SetNoFillComment
\caption{{\ours}RobustPrune%
$(p,\mathcal{C},G,D_p,\alpha,\mathcal{R})$}
\label{alg:grandrobustprune}

     \KwIn{Point $p$; candidate set $\mathcal{C}$ (initially\ containing\ all\ valid nodes); Graph $G$; cached distance map $D_p$, where $D_p[v]=\operatorname{dist}(p,v)$ is the distance between $p$ and $v$ for $v\in\mathcal{C}$; pruning parameter $\alpha$, degree bound $\mathcal{R}$}
\KwOut{$N_{\mathrm{out}}(p)$ updated to contain at most $\mathcal{R}$ nodes}




$N_{\mathrm{out}}(p)\leftarrow\emptyset$\;
\While{$\mathcal{C}\textup{ has valid candidates}$}{
    $p^*\leftarrow\arg\min_{  p'\in\mathcal{C} \wedge (p'\ is\ \text{valid})} D_p[p']$\ ;
    
    $N_{\mathrm{out}}(p)\leftarrow N_{\mathrm{out}}(p)\cup\{p^*\}$\;

    \If{$|N_{\mathrm{out}}(p)|=\mathcal{R}$}{
        \textbf{break}\;
    }

    \ForPar{$p'\in\mathcal{C} \wedge p'\ is$ {$\mathrm{valid}$} }{
        \If{$\alpha\cdot dist(p^*,p')\le D_p[p']$}{
            mark $p'$ as invalid;
        }
    }
}
\end{algorithm2e}

\begin{algorithm2e}[t]
\small 
\DontPrintSemicolon
\SetNoFillComment

   \caption{\ourvamana{}~BatchInsert%
    $(\mathcal{B},G,\ell,\alpha, \mathcal{R}, $P$,k, m)$}
\label{alg:grand-insert}
\SetKwFunction{GS}{BANGSearch}
\SetKwFunction{RP}{GrANDRobustPrune}
\SetKwFunction{CR}{CollectReverseEdges}

\KwIn{batch $\mathcal{B} = \{(p_i,x_i)\}_{i=1}^{B}$, Graph $G$; \beamwidth; pruning factor $\alpha$; degree bound $\mathcal{R}$; scratch capacity $P$; search recall param $k$; graph medoid $m$}
\KwOut{Graph $G$ updated to include all points in $\mathcal{B}$}

\ForPar{$(p_i,x_i)\in\mathcal{B}$}{
 \label{alg:valama-insert-init1}
 write vector $x_i$ to the dataset ; $\deg(p_i)\gets 0$
 \label{alg:valama-insert-init2}
 
}


\ForPar{$(p_i, x_i) \in \mathcal{B}$}{
    ($ V_i,D_{p_i}) \gets$\GS{$G,p_i,k,m, \ell$}\ \tcp{map $D_{p_i}$, containing distance between $p_i$ and $v$ for $v\in V_i$}
  \RP{$p_i,V_i\setminus\{p_i\},G,D_{p_i},
     \alpha,\mathcal{R}$}\;
}

$\mathrm{Dirty} \gets \emptyset$\;
$\mathrm{scratch_x} \gets \emptyset$ for each node $x$ in $G$\;
\coloredtcc{Reverse edges: accumulate into scratch, no locks}

\ForPar{$p_i \in \mathcal{B}$}{
    \ForPar{$u \in N_{\mathrm{out}}(p_i)$}{
        \If{$|\mathrm{scratch}_u| < P$}{$\mathrm{scratch}_u \gets \mathrm{scratch}_u \cup \{p_i\}$\;
        \label{alg:valama-insert-scratch_update}}
        $\mathrm{Dirty} = \mathrm{Dirty} \cup \{u\}$
    }
}

\ForPar{$u \in \mathrm{Dirty}$}{
    $C\gets N_{\mathrm{out}}(u) \cup \mathrm{scratch}_u$\;
    $N_{out}(u) \gets C$ \;

\If{$|\mathcal{C}|>\mathcal{R}$}{
    compute $D_u\gets\operatorname{dist}(u,v)$ for each $v\in\mathcal{C}$ \tcp{map $D_u$, containing distance between $u$ and $v$ for $v\in \mathcal{C}$}
    \RP{$u,\mathcal{C},G,D_u,\alpha,\mathcal{R}$}
}

}
\end{algorithm2e}

\ourvamana implements an optimised version of \textit{RobustPrune}~\cite{diskann}, called \textit{{\ours}RobustPrune} (Algorithm~\ref{alg:grandrobustprune}), which is better suited for parallel processing on GPUs. It avoids redundant distance computation by reusing distances computed during graph traversal.

\noindent\textbf{\ourvamana\ Insert:} Algorithm~\ref{alg:grand-insert} specialises generic batch insertion for Vamana graphs.
Each inserted vector $x_i$ is written to the corresponding slot corresponding to its ID in the dataset, and the corresponding adjacency list of node $p_i$ in the graph is initialised with degree zero. 


The forward neighbourhood of every new node is computed in parallel by using one thread block per node. Each thread block runs \textit{BANGSearch} from a pre-computed medoid to obtain candidates $V_i$, and \textit{{\ours}RobustPrune} selects $\mathcal{R}$ candidates from  $V_i \setminus \{p_i\}$, together costing $\mathcal{O}(\ell \mathcal{R} D)$ work per node, where $D$ is the vector dimensionality. This phase is contention-free because each thread block writes only its own node's adjacency list. Installing the resulting reverse edges is the only phase that touches shared state, since multiple new nodes may select the same existing neighbour $u$. Rather than modifying $N_{\mathrm{out}}(u)$ immediately, \ourvamana\ accumulates every reverse edge $(u,p_i)$ into a per-node scratch buffer (Line~\ref{alg:valama-insert-scratch_update}) of capacity $P$ using an atomic reservation with no locks. Each affected node is added to a \textit{dirty} list. 
Since each new node in the batch adds edges to at most $\mathcal{R}$ vertices, the size of the dirty list is bounded by $B\mathcal{R}$.
Once all reverse edges for the batch have been buffered, the neighbourhood of each dirty node $u$ is rebuilt using one thread block per node. If the size of the new neighbourhood list, $N_{\mathrm{out}}(u)\cup\mathrm{scratch}_u$, is greater than $R$, \textit{{\ours}RobustPrune} is invoked. Processing each node requires $\mathcal{O}((\mathcal{R}{+}P)\mathcal{R}D)$ work. This \emph{accumulate-then-prune} strategy processes a large number of reverse-edge updates using constant-time atomic appends per affected node and at most one prune per affected node per batch, replacing fine-grained lock-protected updates with batched, lock-free reconstruction while preserving the graph quality produced by the original Vamana \textit{RobustPrune}.

\begin{algorithm2e}[t]
  \small \DontPrintSemicolon \SetNoFillComment
   \caption{\ourvamana{}~BatchDelete%
    $(\mathcal{B},G,\ell,\alpha, \mathcal{R}, $P$,k, m, c)$}
  \label{alg:grand-delete}
  \SetKwFunction{GS}{BANGSearch}\SetKwFunction{RP}{GrANDRobustPrune}\SetKwFunction{CC}{ClosestC}\SetKwFunction{MD}{MarkDirty}
  \KwIn{Deletion batch $\mathcal{B} = \{p_i\}_{i=1}^{B}$; graph $G$; \beamwidth; pruning parameter $\alpha$; degree bound $\mathcal{R}$; scratch capacity $P$; search recall param $k$; graph medoid $m$; number of replacement edges $c$} 
  \KwOut{Graph $G$ updated after deleting all points in $\mathcal{B}$}
  \ForPar{$p_i \in \mathcal{B}$}{$(C_i, V_i) \gets$ \GS{$G, p_i, k, m, \ell$}}
  $\mathrm{Dirty} \gets \emptyset$\;
    $\mathrm{scratch_x} \gets \emptyset$ for each node $x$ in $G$\;
  \coloredtcc{In-neighbour repair}
  \ForPar{$p_i \in \mathcal{B}$}{
    \ForPar{$z \in V_i$ \textbf{where} $p_i \in N_{\mathrm{out}}(z)$}{
        $C \gets c$ points of $C_i$ closest to $z$ \;
        $\mathrm{scratch}_z \gets \mathrm{scratch}_z \cup C$ \coloredtcp{size bounded by $P$}
        $\mathrm{Dirty} \gets \mathrm{Dirty} \cup \{z\}$
      }
  }
  \coloredtcc{Out-neighbour repair}
  \ForPar{$p_i \in \mathcal{B}$}{
    \ForPar{$w \in N_{\mathrm{out}}(p_i)$ \textbf{where} $w \not\in \mathrm{B}$}{
        $C \gets c$ points of $C_i$ closest to $w$ \;
        \ForEach{$y \in C$}{$\mathrm{scratch}_y \gets \mathrm{scratch}_y \cup \{w\}$}
        $\mathrm{Dirty} \gets \mathrm{Dirty} \cup C$
    }
  }
  \coloredtcc{Rebuild neighbourhood for dirty nodes}
  \ForPar{$u \in \mathrm{Dirty}$}{
    $C \gets \big(N_{\mathrm{out}}(u) \cup \mathrm{scratch}_u\big) \setminus \mathcal{B}$ \coloredtcp{drops all $p_i$}
    $N_{out}(u) \gets C$ \;
    \If{$|C| > \mathcal{R}$}{
    compute $D_u\gets\operatorname{dist}(u,v)$ for each $v\in\mathcal{C}$ \tcp{map $D_u$, containing distance between $u$ and $v$ for $v\in \mathcal{C}$}
    \RP{$u,\mathcal{C},G,D_v,\alpha,\mathcal{R}$}\;}
  }
\ForPar{$p_i \in \mathcal{B}$}{
    $N_{\mathrm{out}}(p_i) \gets \emptyset$
}
\end{algorithm2e}

\noindent\textbf{\ourvamana\ Delete:} Algorithm~\ref{alg:grand-delete} repairs the graph entirely on the GPU, restoring connectivity around every deleted node in the foreground rather than deferring repair to periodic background consolidation. Unlike IP-DiskANN~\cite{xu2025place}, whose repair is sequential and per-point, every deletion in the batch is repaired concurrently.
For each deleted node $p_i$, a thread-block runs \textit{BANGSearch} in $\mathcal{O}(\ell \mathcal{R} D)$ work to obtain the $C_i$ containing the top-$k$ candidates closest to $p_i$, and a visited set $V_i$ containing vertices whose neighbourhoods may reference $p_i$.

Repair has two stages. \textit{In-neighbor repair} first selects in-neighbors of $p$ from $V_i$. For each in-neighbour $z$, we select $c$ replacement candidates from $C_i$ which are closest to $z$ in $\mathcal{O}(k D)$ work. These candidates are appended to $z$'s scratch buffer, and $z$ is marked dirty. \textit{Out-neighbor repair} does the symmetric operation for each surviving $w \in N_{\mathrm{out}}(p_i)$. We select $c$ replacement candidates from $C_i$ that are closest to $w$, and add $w$ to the scratch buffer of each candidate. The number of replacement edges per repair, $c$, can be tuned to trade repair cost against recall.

The number of dirty vertices added by in-neighbour repair and out-neighbour repair is bounded by $\ell$ and $\mathcal{R}c$ respectively, hence the size of the dirty list is bounded by $B(\ell + \mathcal{R}c)$. Once accumulated, we rebuild the neighbourhood of each dirty node $v$ using a single GPU block per node. If the size of the new neighbourhood list is greater than $\mathcal{R}$, \textit{{\ours}RobustPrune} is invoked. Processing each node requires $\mathcal{O}((\mathcal{R}{+}P)\mathcal{R}D)$ work. After edges are added, the adjacency list of each deleted point $p_i$ is cleared. 

Batched updates, deferred accumulation and lock-free append are techniques that have been used in isolation. \ourvamana combines these to remove two bottlenecks that prior systems have not removed concurrently: (i) multi-index partitioning and background consolidation used by FreshDiskANN, replaced by a single unified graph maintained continuously; and (ii) the need to hold a node's adjacency list under a critical section when modified, which was avoided by having a single owner for each dirty node.

\subsection{\ourcagra}

\begin{algorithm2e}[t]
\small
\DontPrintSemicolon
\SetNoFillComment
  \caption{\ours{}-CAGRA BatchInsert%
    $(\mathcal{B},G,\ell,k)$}
\label{alg:cagra_batch_insert}

\SetKwFunction{CAGRASearch}{CAGRASearch}
\KwIn{batch $\mathcal{B} = \{(p_i,x_i)\}_{i=1}^{B}$, Graph $G$; \beamwidth; search recall param $k$}
\KwOut{Graph $G$ updated to include all points in $\mathcal{B}$}

\ForPar{$(p_i,x_i)\in\mathcal{B}$}{
  write vector $x_i$ to the dataset;   $\deg(p_i)\gets 0$
}
\tcp{Forward-edge creation}
\ForPar{$(p_i, x_i) \in \mathcal{B}$}{
    \coloredtcc{Computed nearest neighbours get into adjacency list  }
    $N_{\mathrm{out}}(p_i) \gets \CAGRASearch(p_i,k, \ell)$
        
}

\tcp{Backward-edge creation}
\ForPar{$p_i\in\mathcal{B}$}{
    \ForPar{$q\in N_{\mathrm{out}}(p_i)$}{
        \tcp{Find detour-edge via $p_i$}
        $C\gets
          N_{\mathrm{out}}(q)\cap N_{\mathrm{out}}(p_i)$\;
        \If{$C\neq\emptyset$}{
        
            select one $r\in C$\;
            atomically replace $r$ with $p_i$ in
            $N_{\mathrm{out}}(q)$\;
        }
    }
}
\end{algorithm2e}

\noindent\textbf{CAGRA Graph Construction:} 
As a first step, CAGRA constructs a proximity graph by generating an approximate k-nearest-neighbour graph using the NN-Descent~\cite{dong2011efficient} algorithm, and then optimises it into a fixed-degree graph. NN-Descent iteratively improves each node’s neighbour list based on the principle that `a neighbour of a neighbour is likely to be a neighbour', enabling efficient graph construction without exhaustive pairwise distance computation. CAGRA and NN-Descent are well-suited to GPUs because graph construction exposes massive parallelism: the neighbour lists of many vertices can be evaluated and updated concurrently. Next, we describe our technique to parallelise insert and delete operations on GPUs.

\noindent\textbf{\ourcagra BatchInsert:} 
To handle dynamic inserts, unlike Vamana, we do not see a need for a heavy pruning procedure along the lines of \textit{RobustPrune}, but instead resort to a lightweight find-and-replace strategy in which accumulating candidate nodes in a buffer attached to the adjacency list is not necessary. We capture the parallelised lock-free algorithm for dynamic insertion on the CAGRA graph in Algorithm~\ref {alg:cagra_batch_insert}. To find the ANNs of the new node to be inserted, we use the existing CAGRA search procedure exposed by the CAGRA library (CAGRASearch). We do not compute more than ${R}$ neighbours and then prune them to fit within the degree bound; we query exactly $\mathcal{R}$ neighbours (i.e. Algorithm~\ref {alg:cagra_batch_insert} is invoked with $k=\mathcal{R}$). This approach intuitively aligns with the CAGRA graph construction policy of the nearest neighbours. Next, for backward edge establishment, we employ a \textit{detour-edge-addition} logic. If the nodes (say $q$) in the adjacency list of $p_i$ have a neighbour that is present in the adjacency list of $p_i$, we replace that neighbour in the adjacency list of $q$ with $p_i$. This replacement can occur without locks, since the last-writer-wins criterion suffices. Overall, this strategy is intuitive, involves minimal distance computations and is lock-free.    











\begin{algorithm2e}[t]
\small
\DontPrintSemicolon
\SetNoFillComment

\caption{\ours-CAGRA BatchDelete$(\mathcal{B},G)$ }
\label{alg:cagra-batch-delete}

\SetKwFunction{PickOne}{PickOne}

\KwIn{Graph $G=(V,E)$,
      deletions $\mathcal{B} = \{p_i\}_{i=1}^{B}$
      }
\KwOut{Graph $G$ updated after deleting all points in $\mathcal{B}$}

\tcp{Compute reverse-neighbour information $N_{in}(z)$}
\ForPar{$v \in V$}{
  $N_{\mathrm{in}}(v)\gets\emptyset$\;
        }

\ForPar{$v \in V$}{
    \ForPar{$z \in N_{\mathrm{out}}(v)$}{
        $N_{\mathrm{in}}(z)
        \gets N_{\mathrm{in}}(z)\cup\{v\}$\;
    }
}
\coloredtcc{$N_{in}(z)$ is stored in CSR format}
\coloredtcc{Rewire IN and OUT nodes of deleted nodes}

\ForPar{$p_i \in \mathcal{B}$}{
    \ForPar{$u \in N_{\mathrm{in}}(p_i) \wedge u\notin\mathcal{B}$}{
        $v \gets \operatorname{PickOne}
        \bigl(N_{\mathrm{out}}(p_i),u\bigr)$\;
        
        $N_{\mathrm{out}}(u)
        \gets
        \bigl(N_{\mathrm{out}}(u)\setminus\{p_i\}\bigr)
        \cup \{v\}$\;
    }
}

\ForPar{$p_i \in \mathcal{B}$}{
    $N_{\mathrm{out}}(p_i) \gets \emptyset$
}

\BlankLine
\SetKwProg{Fn}{Function}{:}{}
\Fn{\PickOne{$L,u$}}{
  \coloredtcc{Pick one node from $L$ using round-robin policy}
$r\gets\text{an element of $L$ such that }
r\notin\mathcal{B}\land r\neq u
\land r\notin N_{\mathrm{out}}(u)$\;
\Return{r}\;

}
    
\end{algorithm2e}
\noindent\textbf{\ourcagra BatchDelete:} 
To handle deletions, finding the incoming edges is the key; there is no reliable way to trace them. The naive approach would be to scan the entire graph for the IN nodes of $p_i$, and this would be prohibitive as we scale the dataset. We experimented with other logical alternatives, such as BFS or a scan of a limited region around $p_i$, but we were unable to reliably recover all IN nodes. Finally, we implement a novel and reliable approach to build a reverse graph index in the regular (forward) index graph. The main challenge with this reverse graph is the increased memory utilisation (especially on a GPU with limited memory), because the reverse graph corresponding to a regular graph need not be regular.

To overcome this, we implement a reverse graph in the Compressed Sparse Row (CSR) format. This would require computing the reverse graph in-place each time the delete batch is processed. Empirically, we find that reverse graph computation on a GPU is an order of magnitude faster than a naive linear-scan-based comparison for detecting whether $p_i$ occurs across the entire graph. To complement this, the CAGRA graph (NN-descent style) includes an additional reverse graph in approximately the same space as the Vamana graph. We present the algorithm in Algorithm~\ref{alg:cagra-batch-delete}.  To replace the OUT node in the IN node's adjacency list, we heuristically select a node from the OUT set in a round-robin manner.
\section{Experimental Setup}
\label{s:exp_setup}

\subsection{Datasets}

We evaluate \ours on seven widely used datasets with diverse scales and data types. For each dataset, we use subsets of sizes 1 million, 10 million, and 100 million points, depending on the availability of the particular size and the amount of GPU memory available to accommodate the dataset and the graph index. Table \ref{tab2} lists the specific datasets used for evaluation.


\begin{table}[htbp]
\centering
\small
\setlength{\tabcolsep}{3pt}      

\begin{tabular}{|l|c|@{~}c@{~}c@{~}c@{~}|@{~}c@{~}|@{~}c@{~}|@{~}c@{~}|}
\hline
\textbf{Dataset} &
\textbf{Full} &
\multicolumn{3}{@{~}c|@{~}}{\textbf{Sizes Used}} &
\textbf{Query} &
\textbf{Distance} &
\textbf{Vector} \\
&
\textbf{Size} &
\textbf{1M} &
\textbf{10M} &
\textbf{100M} &
\textbf{Size} &
\textbf{Metric} &
\textbf{Dim.}
\\
\hline
GloVe-100  & 1.2M & Y & N & N & 10,000 & Cosine     & 100 \\
Wikipedia  & 35M  & Y & Y & N & 5,000  & IP         & 768 \\
MSMARCO    & 100M & Y & Y & N & 9,376  & IP         & 768 \\
Text2Image & 1B  & Y & Y & N & 10,000 & IP         & 200 \\
MSTuring   & 1B   & Y & Y & Y & 10,000 & Euclidean  & 100 \\
Deep-1B    & 1B   & Y & Y & Y & 10,000 & Euclidean  & 96 \\
SIFT1B     & 1B   & Y & Y & Y & 10,000 & Euclidean  & 128 \\
\hline
\end{tabular}
\caption{Datasets used for evaluation.}
\label{tab2}
\end{table}

\subsection{Workload Generation}

\begin{figure}
    \centering
    \includegraphics[width=0.9\linewidth]{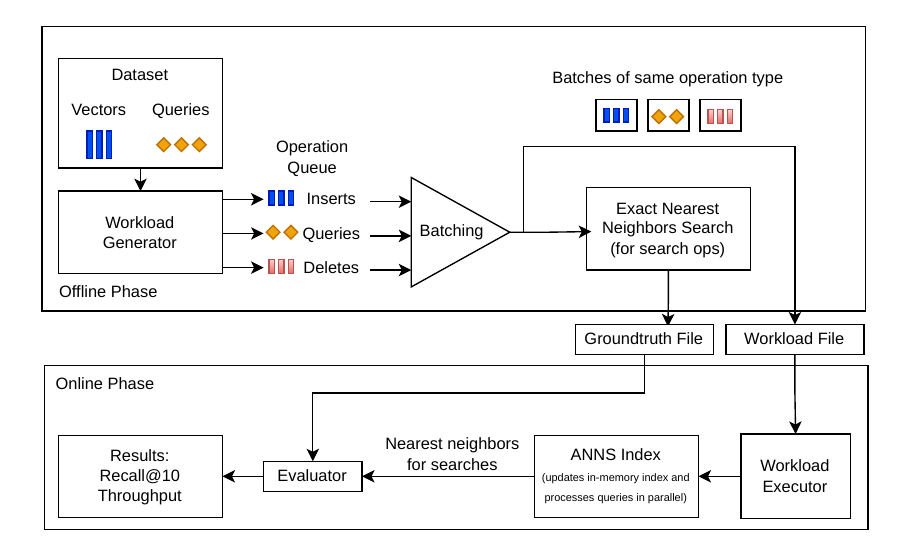}
    \caption{Streaming Workload Generation Pipeline}
    \label{fig:wlg-pipeline}
\end{figure}

Dynamic ANNS systems are subjected to a sequence of insert, delete, and search operations interleaved over time. Each search operation has a ground-truth corresponding to the set of active points at the time the search is issued. We implement a dedicated streaming workload generator for our evaluation, as shown in Figure~\ref{fig:wlg-pipeline}. For each dataset, the workload generator (i) loads the base and query vectors, (ii) generates an operation stream of batches following streaming workload patterns, and (iii) computes the exact top-100 nearest neighbours for each query against the current active set at each search step. Batches are of uniform sizes,  consist of operations of a single type and are issued sequentially. The streaming workload patterns are based on ones from CleanANN~\cite{zhang_cleanann_2026}, Quake~\cite{mohoney_quake_2025} and the Streaming Track of the NeurIPS'23 Big-ANN Competition~\cite{simhadri2026results}.

For each dataset, we generate streaming workload patterns which consist of insert, delete, and query operations. These workloads mimic real-world access patterns and vary in the spatial correlation between vectors inserted/deleted in a single step.

\noindent \textbf{Sliding Window}: The dataset is divided into 100 segments, each containing an equal number of points, and the first 50 segments are active at the start. On each timestep $t$, the window is shifted by first inserting the $(t + 50)^{th}$ segment, and then deleting the $t^{th}$ segment. A search operation is issued after both insert and delete.

\noindent \textbf{Interleaved}: This workload models random inserts and deletions. Each operation is generated randomly following a pre-configured distribution (50\% Insert, 30\% Search, 20\% Delete). For each delete operation, a batch of vectors is randomly selected from the active set and deleted. Operations are generated until there are no more vectors to be inserted.

\noindent \textbf{Expiration Time}: Each point in the dataset is assigned different lifetimes: short-term (10 cycles), long-term (100 cycles) and permanent in a 10:2:1 ratio. The second half of the dataset is divided into 50 segments. In each cycle, one segment of points is inserted, and all the points which expire in that cycle are deleted. A search operation is issued after both the insert and the delete.

\noindent \textbf{Clustered}: The points in the dataset are partitioned into 64 clusters using K-means clustering. Then, the clusters are processed in a sliding window fashion. The first 32 clusters are inserted at the start, and on the $t^{th}$ timestep, the $(t+32)^{th}$ cluster is inserted, and the $t^{th}$ cluster is deleted. Search operations are issued during the insertion and deletion of each cluster whenever 1\% of the dataset has been inserted/deleted.

\noindent \textbf{Insert Heavy}: The dataset grows to full capacity, and operations are issued with a ratio of 90\% insertions to 10\% searches. No deletes are performed. This lets us evaluate algorithms when the number of active points grows quickly. 

\subsection{Baselines}

We evaluate \ours against two state-of-the-art Streaming ANNS systems: SVFusion and \ourfda.

\noindent \textbf{SVFusion} \cite{peng_svfusion_2026} is a state‑of‑the‑art GPU streaming
index we take as the reference for the CAGRA graph family. It is a CPU-GPU multi-tier index, which uses both the CPU and the GPU for data storage. It includes a delete‑consolidation subsystem that rebuilds affected graph regions after deletions.

\noindent \textbf{FreshDiskANN} \cite{freshdiskann} is a state-of-the-art graph-based ANNS index that supports dynamic workloads in memory or on SSDs. There is no public GPU implementation of FreshDiskANN, so we implement one from scratch to serve as the baseline for Vamana graphs.

Our implementation, \textbf{\ourfda}, is a faithful GPU port of FreshDiskANN.
A detailed explanation of the algorithms is available in the supplementary material.
We maintain three index types: a Long-Term Index (LTI) containing the consolidated base graph, $M$ read-only temporary indexes holding up to $T$ points each, and a read-write index with capacity $T$ for insertions. Locks are only required for the read-write index, as other indexes are read-only.

Points are inserted into the read-write index by performing \textit{BANGSearch} with beam width $\ell$ followed by \textit{{\ours}RobustPrune} with parameter $\alpha$. 
Deletions use a delete bitset of size $D$, and when the bitset reaches capacity, a background \textbf{merge} operation is triggered to remove tombstoned points and repair neighbourhoods.

Once the read-write index is full, it is sealed and becomes a read-only temporary index. When the maximum number of temporary indexes is reached, the background merge operation consolidates them into the LTI index while removing deleted points. Each index functions as an independent subgraph with its own medoid. Search queries perform a greedy search over each index independently, and pick the overall top-$k$ nearest neighbours.

\subsection{Algorithm Parameters}

\begin{figure*}
    \centering
    \includegraphics[width=\linewidth]{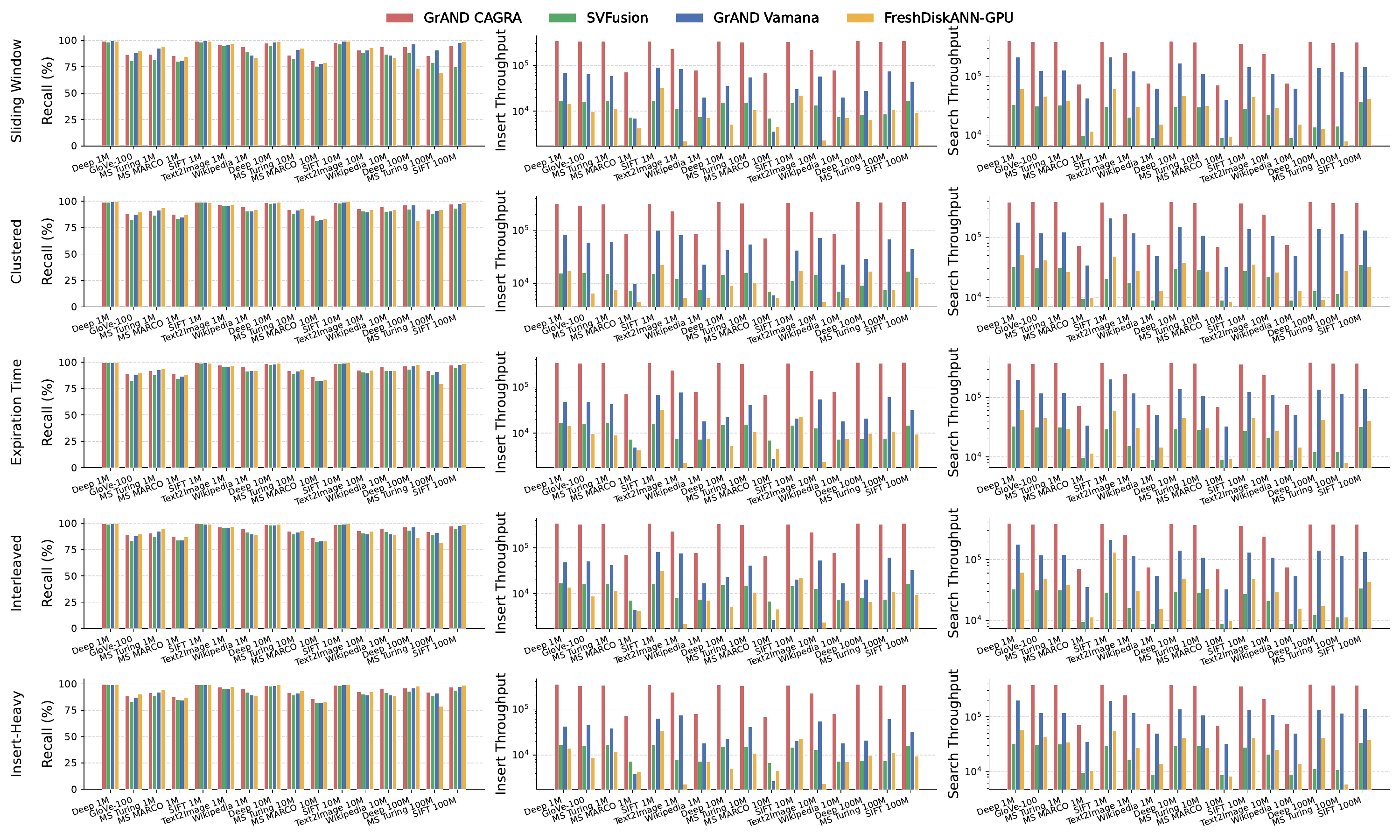}
    \caption{Average recall, insert throughput and search throughput for various workload patterns}
    \label{fig:overall_results}
\end{figure*}

To ensure that the evaluation is fair, we ensure that all the systems consume the same operation stream and are scored against the same top-100 ground truth. We also ensure that all seed graphs are built with a consistent seed.  For datasets that use the inner product distance metric, we augment MIPS distance into L2 distance by adding an extra dimension $x \to [x, \sqrt{M^2 - \Vert x\Vert^2}]$, where $M$ is the maximum base norm.

For 100M datasets, the delete-consolidation performed by SVFusion, and the merge operation performed by \ourfda exceed memory limits on the GPU, so we run these systems with consolidation disabled, making deletes tombstone‑only.

\ourvamana is run with beam size $\ell = 100$, pruning parameter $\alpha = 1.2$, number of scratch slots $P = 64$, recall param $k=50$, and number of replacement edges $c = 2$. 
\ourfda is run with insert beam width $\ell = 100$, pruning parameter $\alpha = 1.2$, number of temporary indexes $M = 4$, and delete list capacity $D = 2,000,000$. The capacity of each temporary index is set to the number of points to be inserted, divided by the number of temporary indexes, $M$.

\ourcagra is run with graph degree $\mathcal{R} = 32$, intermediate graph degree 64, and search itopk size 256. SVFusion is run with graph degree $\mathcal{R} = 64$, intermediate graph degree 128, and search itopk size 256.

\subsection{Metrics}

We focus on two key metrics for evaluation:
\noindent \textbf{Recall}: Recall@k is the fraction of $k$ actual nearest neighbours among the top $k$ returned by the algorithm. We use $k = 10$.
\noindent \textbf{Throughput}: We measure throughput for each operation type (insert, search, delete) as the number of points processed per unit time.



\section{Evaluation}
\label{s:eval}

We evaluate \ours using both the Vamana and CAGRA graph families under a variety of streaming workloads and dataset scales. Our evaluation seeks to answer the following questions:

\begin{itemize}
    \item Does \ours improve insert, delete, and search throughput compared to existing dynamic graph indexes?
    \item How does \ours perform under different streaming workloads and dataset scales?
    \item Does \ours sustain the high search accuracy and maintain high throughput during long-running streaming execution?
\end{itemize}

The remainder of this section answers these questions through experiments on seven benchmark datasets using five representative streaming workloads.

\subsection{Average Performance over Workloads}
\label{sec:workloads}

\begin{figure*}
    \centering
    \includegraphics[width=\linewidth]{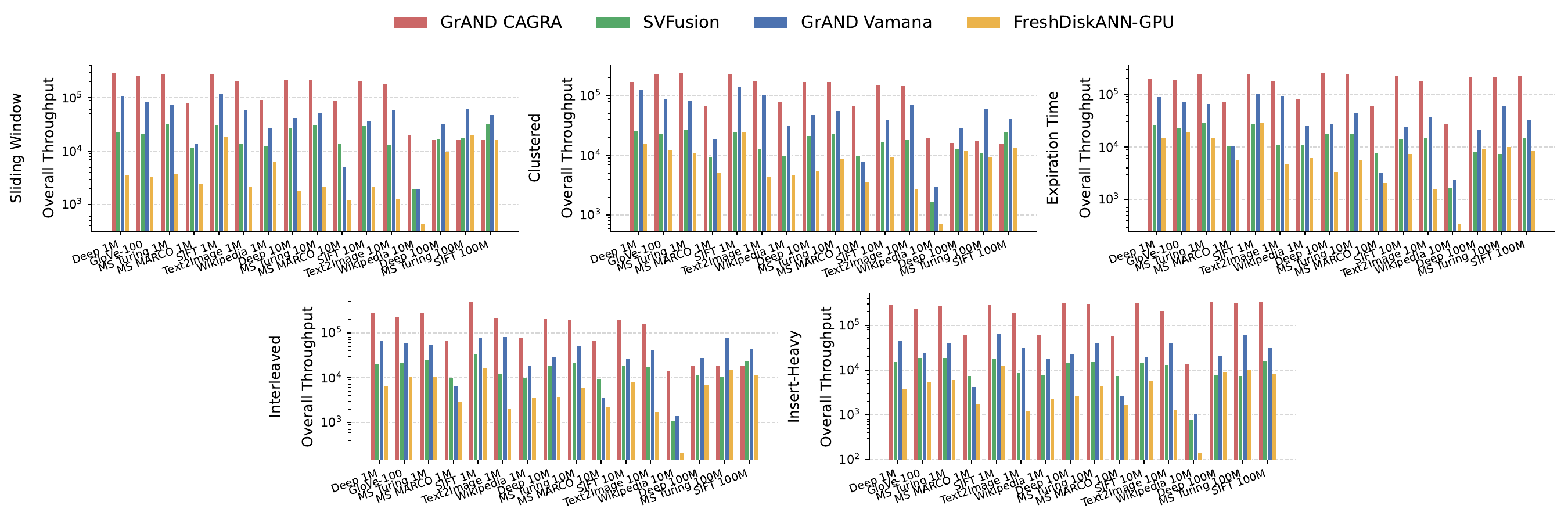}
    \caption{Overall throughput for various workload patterns}
    \label{fig:overall_throughput}
\end{figure*}

Figure \ref{fig:overall_results} shows the average recall, insert throughput, and search throughput for various workload patterns on each dataset.

\ourvamana and \ourcagra achieve recall comparable to both baselines while significantly outperforming them in throughput. \ourvamana achieves an insert throughput $5.09\times$ higher  (geomean)  compared to \ourfda and $2.59\times$ higher on average compared to SVFusion. Similarly, \ourcagra achieves an insert throughput that is $39.89\times$ higher on average than \ourfda and $20.29\times$ higher on average than SVFusion. Similarly, for search throughput, \ourvamana achieves a search throughput $5.64\times$ higher on average than \ourfda and $5.51\times$ higher on average than SVFusion. \ourcagra achieves a search throughput that is $13.79\times$ higher on average than \ourfda and $13.47\times$ higher on average than SVFusion.

It is not possible to directly measure delete throughput for SVFusion and \ourfda as they use lazy deletion. Delete operations are processed quickly, but the actual graph repair happens in the background. To account for this, we also measure the overall throughput for each index.

Figure~\ref{fig:overall_throughput} shows the overall throughput for each workload type on various datasets. \ourvamana achieves $6.50\times$ higher overall throughput than \ourfda and $2.24\times$ higher overall throughput than SVFusion on average. Similarly \ourcagra achieves overall throughput $25.42\times$ higher than \ourfda and $8.76\times$ higher than SVFusion on average.

\begin{figure*}[h]
    \centering
\includegraphics[width=0.9\linewidth]{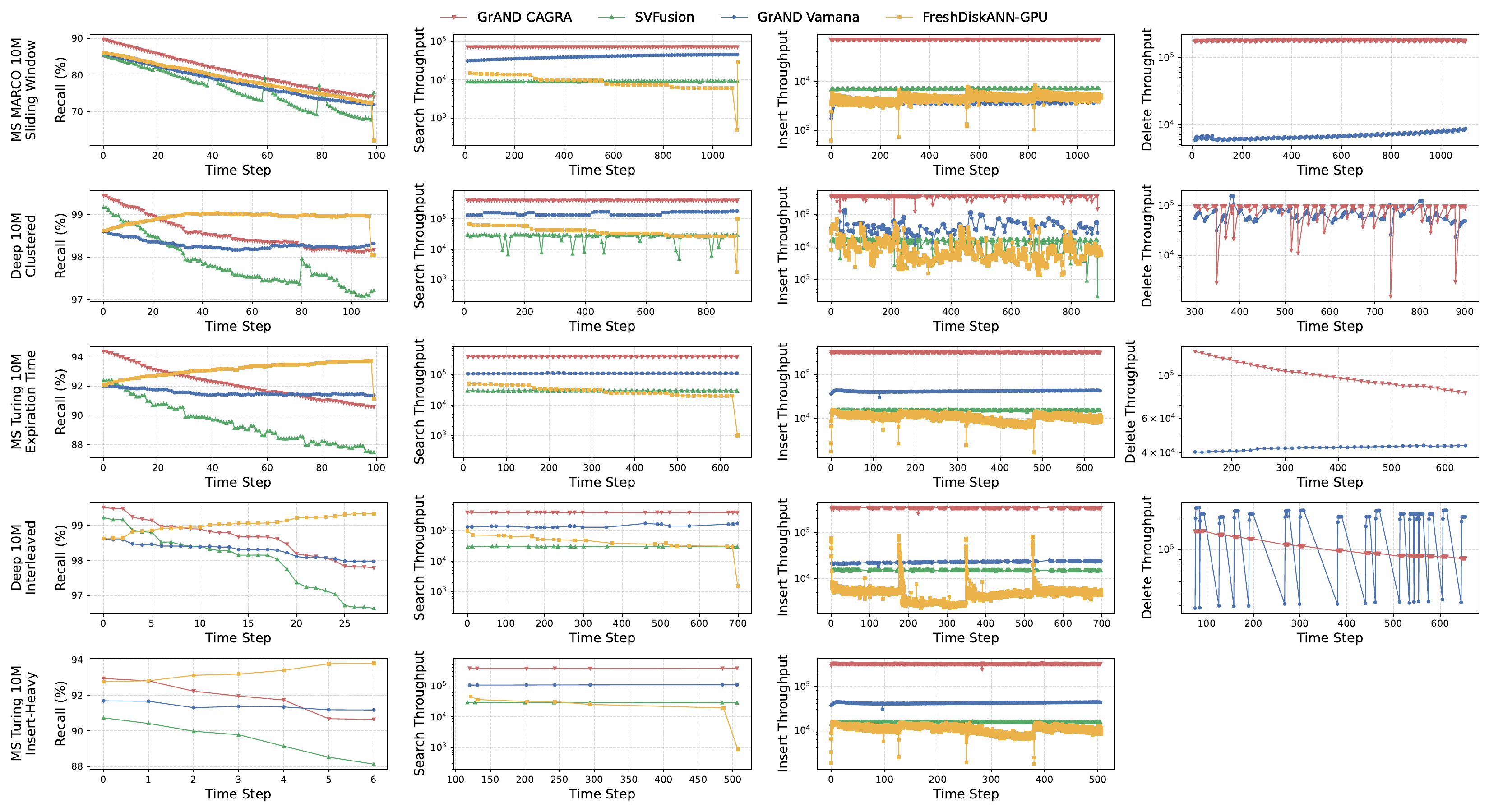}
    \caption{Recall and throughputs over time on selected 10M workloads}
    \label{fig:10m_results}
\end{figure*}

For workloads on 100M datasets, \ourfda and SVFusion were run with consolidation disabled due to memory constraints on the GPU. As tombstoned vertices accumulate, this leads to a decrease in recall. This can be observed in the MSTuring 100M Sliding Window workload, where SVFusion's recall fell by $7\%$ compared to \ourcagra, and \ourfda's recall fell by $21\%$ compared to \ourvamana. This recall drop would be greater on longer and more delete-heavy workloads.

Additionally, on 100M workloads, \ourcagra also exhibits a reduction in overall throughput. This is primarily due to the additional cost of repairing neighbourhoods affected by deletion.
If consolidation were enabled for \ourfda and SVFusion, the additional consolidation overhead would reduce overall throughput. Therefore, the reported throughput values do not represent a direct comparison of performance under equivalent settings.

We observe degradation of insert throughput for \ourvamana on the MSMARCO dataset. \ourvamana outperforms SVFusion on the Wikipedia dataset but not on MSMARCO, even though the two datasets have the same dimensionality, distance metric, and embedding type (i.e. text). \ourfda and \ourvamana, both of which are based on the Vamana graph, show approximately $4\times$ reduction in insert throughput on MSMARCO as compared to Wikipedia, while insert throughputs of SVFusion and \ourcagra are not drastically affected. Our analysis suggests that this difference is because MSMARCO is less clustered than Wikipedia, resulting in a more dispersed graph, which leads to nodes having higher degrees and a larger number of candidates for reverse-edge pruning. We observed that inserting a batch of points into MSMARCO required adding reverse edges to approximately $4\times$ more vertices than that of Wikipedia, leading to the observed reduction in insertion throughput.

We now compare the performance for various workload types:

\noindent \textbf{Sliding Window}: The sliding window workload involves gradual replacement of data, allowing all systems to maintain steady throughput throughout the workload. We observe occasional drops in the insert throughput for SVFusion and \ourfda, which can be attributed to background consolidation. However, such drops do not occur for \ourvamana or \ourcagra.

\noindent \textbf{Clustered}: In the clustered workload, consecutive insertions and deletions affect neighbouring vertices. The inserts and deletes are localised in nature, which causes significant variations in throughput. \ourfda experiences increased synchronisation overhead on insert operations due to repeated reverse-edge updates and neighbourhood pruning, which is avoided in \ourvamana by the accumulation of reverse-edge updates.

\noindent \textbf{Expiration Time}: In expiration time workloads, the change in the dataset is gradual, but more erratic than sliding window workloads. The throughput remains stable, but we can see recall drops in SVFusion and \ourfda due to tombstone-based deletion, which do not occur in \ourvamana or \ourcagra.

\noindent \textbf{Interleaved}: The interleaved workload deletes random points, which highlights the need for dynamic graph repair. \ourfda incurs overheads from immediate reverse-edge updates and periodic index management, while SVFusion faces overheads due to CPU-GPU transfer latency. \ours maintains high throughput under this workload due to local neighbourhood repair and the accumulation of reverse edges.

\noindent \textbf{Insert-Heavy}: On insert-heavy workloads, \ourfda incurs overhead from maintaining multiple graph partitions, index transitions, and periodic merge operations. In contrast, \ours inserts directly into a unified graph with incremental GPU-based repairs, avoiding graph migration and consolidation.

\subsection{Temporal Performance on Workloads}
\label{sec:10m}
On streaming workloads, consistent throughput is just as important as average throughput. To analyse the performance over time, we evaluate each method on datasets containing up to 10 million (10M) vectors, which provide sufficient scale to expose the overheads of dynamic graph maintenance while allowing all systems to operate under their intended configurations.
Figure~\ref{fig:10m_results} shows the recall and throughputs for inserts, searches, and deletes, achieved by each method for selected workloads. Delete throughput is omitted for Insert-Heavy workloads because they do not include delete operations.
We analyse recall, insertion, deletion and search separately.

\noindent \textbf{Recall}:
Both \ours implementations maintain recall comparable to their respective baselines across all workloads while maintaining significantly higher update throughput. This is expected, since both \ourvamana and \ourcagra use the same search algorithm as \ourfda and SVFusion, respectively, but differ in how they apply graph updates.

We observe periodic drops in recall in \ourfda and SVFusion. These fluctuations arise from tombstone-based deletes. At larger scales, where consolidation becomes infrequent or is disabled, stale edges accumulate and recall gradually degrades under delete-heavy workloads, whereas \ours continues to preserve graph quality through online maintenance.

\noindent \textbf{Search Throughput}:
Both \ours implementations achieve higher search throughput while maintaining recall comparable to the baselines. \ourvamana achieves higher search throughput than \ourfda by maintaining a unified graph rather than searching across multiple graph partitions. Similarly, \ourcagra\ avoids the CPU-GPU coordination required by SVFusion by executing searches entirely on the GPU. Search throughput decreases  as distance computations become more expensive, but the relative performance trends remain consistent.

\noindent \textbf{Insert Throughput}:
\ourvamana and \ourcagra consistently achieve higher insert throughput than the corresponding baselines. For \ourvamana, the improvement primarily stems from batching reverse-edge updates, which substantially reduces synchronisation overhead.
Similarly, \ourcagra\ consistently outperforms SVFusion by performing graph maintenance directly on the GPU and eliminating the need for CPU-GPU coordination.

We observe fluctuations in insert throughput for both \ourfda and SVFusion. This can be attributed to the overhead of background graph repair. In contrast, \ours maintains a single continuously updated graph. By avoiding temporary indices and global consolidation, the amount of work performed per batch remains consistent over time, providing stable throughput even under long-running streaming workloads. In SVFusion, because graph metadata resides on the CPU while graph traversal runs on the GPU, fluctuation in host-device synchronisation also affects overall throughput. 

\noindent \textbf{Delete Throughput}:
\ourfda and SVFusion use lazy deletion, in which deleting a point only marks it with a tombstone and defers graph repair until a later consolidation phase. Hence, a delete operation can be processed almost instantly and does not reflect the deferred cost of maintaining the graph. 
Hence, it becomes difficult to compute delete throughputs for these systems.

\ourvamana and \ourcagra use eager in-place deletion. \ourcagra generally achieves higher delete throughput (up to $20\times$ faster) than \ourvamana as it avoids expensive adjacency list pruning. For clustered workloads, the localised nature of deletes allows \ourvamana to prune adjacency lists more quickly and achieve delete throughput comparable to \ourcagra. However, for 100M datasets, the cost of computing in-neighbours dominates, and \ourcagra experiences a drop in delete throughput. Under interleaved workloads, \ourvamana\ experiences periodic drops in delete throughput when a batch removes the graph's medoid. Deleting the medoid requires installation of a new medoid, incurring a performance penalty.

\section{Related Work}
\label{s:related}


Graph-based ANNS has become the dominant approach for large-scale vector retrieval due to its balance between search accuracy and query latency. As modern vector-based applications rapidly evolve, maintaining the underlying graph indexes amid continuous insertions and deletions has become a fundamental systems challenge. We categorise existing ANN search into four categories: static index, CPU-based dynamic index, CPU-GPU hybrid dynamic index and GPU-based dynamic index.

\noindent\textbf{Static Index.} Graph-based ANNS methods construct navigable proximity graphs that enable efficient approximate nearest neighbour search through greedy graph traversal. Early systems such as HNSW~\cite{malkov_efficient_2018} and NSG~\cite{fu_fast_2017} demonstrated that carefully designed neighbourhood structures provide high recall while maintaining logarithmic search complexity. More recently, Vamana~\cite{diskann} introduced a sparse navigable graph optimised for billion-scale datasets, forming the foundation of DiskANN and several subsequent graph-based ANN systems. ParlayANN~\cite{10.1145/3627535.3638475} develops parallel implementations for Vamana graphs.

GPU-accelerated ANN search, e.g.  FAISS~\cite{johnson_billion-scale_2019}, SONG~\cite{zhao_song_2020}, GANNS~\cite{yu_gpu-accelerated_2022}, BANG~\cite{BANG}, FusionANNS~\cite{tian_fusionanns_2024}, PilotANN~\cite{Gui2026}, G\textsc{ust}ANN~\cite{Jiang2025} and CAGRA~\cite{ootomo2024cagra} deliver significantly higher throughputs over pure CPU-based methods~\cite{liu2026gpu}. These systems assume that the graph index is constructed offline and remains fixed during query processing, limiting their applicability to continuously evolving datasets. GrAND builds on the search and construction primitives of Vamana and CAGRA, but replaces their static assumption with support for concurrent, in-place updates entirely on the GPU.

\noindent\textbf{CPU-Based Dynamic Index.} FreshDiskANN~\cite{freshdiskann} extends the DiskANN index with incremental updates by maintaining newly inserted points in auxiliary graph structures and periodically consolidating them into a long-term index, and deleted points are flagged for lazy deletion. This design reduces rebuild costs, but periodic consolidation introduces additional maintenance overhead, and graph quality depends on the consolidation frequency. 
IP-DiskANN~\cite{xu2025place} is the first work to propose an in-place deletion strategy that is applicable to Vamana graphs and avoids periodic consolidation overhead. 
CleanANN~\cite{zhang_cleanann_2026} similarly targets fully dynamic graph-based ANNS on CPUs, combining workload-aware neighbour linking to address data distribution shift with lock-free, semi-lazy memory cleaning to bound the overhead of deletion-induced graph repair. Outside the graph-based setting, SPFresh~\cite{xu_spfresh_2023} performs incremental in-place updates on a disk-resident, partition-based (IVF) index with lightweight rebalancing, while Quake~\cite{mohoney_quake_2025} similarly adopts a multi-level partitioned index, using a cost model to guide adaptive partition splits and merges and to tune query parameters as the index evolves. O\textsc{DIN}ANN~\cite{Guo2026} directly inserts vectors to the disk instead of buffering them in memory. In contrast to these CPU-based systems, GrAND performs repair entirely on the GPU using massively parallel, lock-free batch updates rather than per-update locking or background consolidation, avoiding the serialisation and host-side coordination these systems incur.

\noindent\textbf{CPU-GPU Hybrid Dynamic Index.} 
Implementing streaming ANNS is computationally intensive, and hence CPU-based techniques suffer from low throughput. 
SVFusion~\cite{peng_svfusion_2026} introduces a collaborative CPU-GPU-SSD architecture for streaming vector search, combining hierarchical vector storage, workload-aware caching, and cooperative execution across all three tiers to improve scalability under dynamic workloads. This hierarchical storage management, along with periodic consolidation across tiers, introduces additional coordination overhead.
To avoid data transfer and synchronisation overheads between CPU/GPU, GrAND targets high throughput for datasets that fit entirely within GPU memory.

\noindent\textbf{GPU-Based Dynamic Index.} A relatively smaller body of work targets dynamic updates on GPU-resident indexes. Jasper~\cite{jasper} supports incremental index construction via batched inserts on a GPU-resident Vamana graph, combining lock-free batch-parallel insertions, GPU-efficient graph traversal, and RaBitQ quantization to sustain high search throughput. However, it does not support streaming vector deletions. SIVF~\cite{zhao_sivf_2026} similarly targets GPU-resident mutability, but for an IVF rather than a graph-based index, introducing conflict-free slab allocation and coalesced search over non-contiguous GPU memory to enable low-latency in-place updates within Faiss. GrAND differs from both by performing in-place deletion with immediate neighbourhood repair on graph indexes, and by evaluating this design across two distinct graph families (Vamana and CAGRA) under a common batched, lock-free execution model, without relying on vector compression.




\section{Conclusion}
We presented \ours, a set of GPU-native high-performance algorithms for maintaining dynamic graph-based ANNS indexes under streaming workloads, enabling searches, insertions, and deletions to operate directly without periodic index rebuilds or CPU-assisted synchronisation. Its design combines batched and lock-free graph updates with cumulative pruning to effectively eliminate redundant computation, efficient resource-pool memory management, and a GPU-friendly in-place deletion strategy that accurately identifies and repairs deletion-affected edges. These techniques are general enough to support distinct graph structures, as demonstrated through their integration with both Vamana and CAGRA. 
\ours achieved $2.2\times$--$8.7\times$ higher average throughput than competing approaches.



\newpage



\bibliographystyle{ACM-Reference-Format}
\bibliography{main}

@inproceedings{diskann,
 author = {Jayaram Subramanya, Suhas and Devvrit, Fnu and Simhadri, Harsha Vardhan and Krishnawamy, Ravishankar and Kadekodi, Rohan},
 booktitle = {Advances in Neural Information Processing Systems},
 editor = {H. Wallach and H. Larochelle and A. Beygelzimer and F. d\textquotesingle Alch\'{e}-Buc and E. Fox and R. Garnett},
 pages = {},
 publisher = {Curran Associates, Inc.},
 title = {DiskANN: Fast Accurate Billion-point Nearest Neighbor Search on a Single Node},
 volume = {32},
 year = {2019}
}

@article{cuda,
title = {CUDA: Compiling and optimizing for a GPU platform},
journal = {Procedia Computer Science},
volume = {9},
pages = {1910-1919},
year = {2012},
note = {Proceedings of the International Conference on Computational Science, ICCS 2012},
issn = {1877-0509},
doi = {https://doi.org/10.1016/j.procs.2012.04.209},
url = {https://www.sciencedirect.com/science/article/pii/S1877050912003304},
author = {Gautam Chakrabarti and Vinod Grover and Bastiaan Aarts and Xiangyun Kong and Manjunath Kudlur and Yuan Lin and Jaydeep Marathe and Mike Murphy and Jian-Zhong Wang}
}

@misc{cudablog,
  title = {{CUDA Programming Model}},
  howpublished = {\url{https://docs.nvidia.com/cuda/pdf/CUDA_C_Programming_Guide.pdf}},
  author = {CUDA},
  year = {2025},
}

@ARTICLE{BANG,
author={Venkatasubba, Karthik and Khan, Saim and Singh, Somesh and Simhadri, Harsha Vardhan and Vedurada, Jyothi},
journal={ IEEE Transactions on Big Data },
title={{ BANG: Billion-Scale Approximate Nearest Neighbour Search Using a Single GPU }},
year={2025},
volume={11},
number={06},
ISSN={2332-7790},
pages={3142-3157},
doi={10.1109/TBDATA.2025.3581085},
url = {https://doi.ieeecomputersociety.org/10.1109/TBDATA.2025.3581085},
publisher={IEEE Computer Society},
address={Los Alamitos, CA, USA},
month=dec}

@InProceedings{Sun2024,
  author     = {Sun, Yiping and Shi, Yang and Du, Jiaolong},
  booktitle  = {Proceedings of the 33rd ACM International Conference on Information and Knowledge Management},
  title      = {A Real-Time Adaptive Multi-Stream GPU System for Online Approximate Nearest Neighborhood Search},
  year       = {2024},
  pages      = {4906--4913},
  readstatus = {skimmed},
}

@inproceedings{zhang_cleanann_2026,
author = {Zhang, Ziyu and Wei, Yuanhao and Engels, Joshua and Shun, Julian},
title = {CleanANN: Efficient and Robust Full Dynamism in Graph-based Approximate Nearest Neighbor Search},
year = {2026},
isbn = {9798400727610},
publisher = {Association for Computing Machinery},
address = {New York, NY, USA},
url = {https://doi.org/10.1145/3816782.3819219},
doi = {10.1145/3816782.3819219},
booktitle = {Proceedings of the 38th ACM Symposium on Parallelism in Algorithms and Architectures},
pages = {247–260},
numpages = {14},
location = {Royal Holloway, University of London, London, United Kingdom},
series = {SPAA '26}
}

@article{zhao_song_2020,
	title = {{SONG}: {Approximate} nearest neighbor search on {GPU}},
	booktitle = {2020 {IEEE} 36th {International} {Conference} on {Data} {Engineering} ({ICDE})},
	publisher = {IEEE},
	author = {Zhao, Weijie and Tan, Shulong and Li, Ping},
	year = {2020},
	pages = {1033--1044},
}

@article{johnson_billion-scale_2019,
	title = {Billion-scale similarity search with gpus},
	volume = {7},
	number = {3},
	journal = {IEEE Transactions on Big Data},
	author = {Johnson, Jeff and Douze, Matthijs and Jégou, Hervé},
	year = {2019},
	note = {Publisher: IEEE},
	pages = {535--547},
}

@article{fu_fast_2017,
	title = {Fast approximate nearest neighbor search with the navigating spreading-out graph},
	journal = {arXiv preprint arXiv:1707.00143},
	author = {Fu, Cong and Xiang, Chao and Wang, Changxu and Cai, Deng},
	year = {2017},
}

@article{malkov_efficient_2018,
	title = {Efficient and robust approximate nearest neighbor search using hierarchical navigable small world graphs},
	volume = {42},
	number = {4},
	journal = {IEEE transactions on pattern analysis and machine intelligence},
	author = {Malkov, Yu A and Yashunin, Dmitry A},
	year = {2018},
	note = {Publisher: IEEE},
	pages = {824--836},
}

@inproceedings{yu_gpu-accelerated_2022,
	title = {{GPU}-accelerated {Proximity} {Graph} {Approximate} {Nearest} {Neighbor} {Search} and {Construction}},
	doi = {10.1109/ICDE53745.2022.00046},
	booktitle = {2022 {IEEE} 38th {International} {Conference} on {Data} {Engineering} ({ICDE})},
	author = {Yu, Yuanhang and Wen, Dong and Zhang, Ying and Qin, Lu and Zhang, Wenjie and Lin, Xuemin},
	year = {2022},
	pages = {552--564},
}

@article{tian_fusionanns_2024,
	title = {{FusionANNS}: {An} {Efficient} {CPU}/{GPU} {Cooperative} {Processing} {Architecture} for {Billion}-scale {Approximate} {Nearest} {Neighbor} {Search}},
	journal = {arXiv preprint arXiv:2409.16576},
	author = {Tian, Bing and Liu, Haikun and Tang, Yuhang and Xiao, Shihai and Duan, Zhuohui and Liao, Xiaofei and Zhang, Xuecang and Zhu, Junhua and Zhang, Yu},
	year = {2024},
}

@article{freshdiskann,
  author  = {Singh, Aditi and Jayaram Subramanya, Suhas and Krishnaswamy, Ravishankar and Simhadri, Harsha Vardhan},
  journal = {arXiv preprint arXiv:2105.09613},
  title   = {{FreshDiskANN}: A Fast and Accurate Graph-Based {ANN} Index for Streaming Similarity Search},
  year    = {2021},
}

@article{jasper,
  author  = {McCoy, Hunter and Wang, Zikun and Pandey, Prashant},
  title   = {{GPU}-Accelerated {ANNS}: Quantized for Speed, Built for Change},
  journal = {arXiv preprint arXiv:2601.07048},
  year    = {2026},
}

@article{peng_svfusion_2026,
  title   = {{SVFusion}: A {CPU}-{GPU} Co-Processing Architecture for Large-Scale Real-Time Vector Search},
  author  = {Peng, Yuchen and Yang, Dingyu and Xie, Zhongle and Sun, Ji and Shou, Lidan and Chen, Ke and Chen, Gang},
  journal = {Proceedings of the VLDB Endowment (PVLDB)},
  volume  = {19},
  year    = {2026},
  note    = {Accepted for VLDB 2026; arXiv preprint arXiv:2601.08528},
}

@inproceedings{ootomo2024cagra,
  title={Cagra: Highly parallel graph construction and approximate nearest neighbor search for gpus},
  author={Ootomo, Hiroyuki and Naruse, Akira and Nolet, Corey and Wang, Ray and Feher, Tamas and Wang, Yong},
  booktitle={2024 IEEE 40th International Conference on Data Engineering (ICDE)},
  pages={4236--4247},
  year={2024},
  organization={IEEE}
}

@inproceedings{dong2011efficient,
  title={Efficient k-nearest neighbor graph construction for generic similarity measures},
  author={Dong, Wei and Moses, Charikar and Li, Kai},
  booktitle={Proceedings of the 20th international conference on World wide web},
  pages={577--586},
  year={2011}
}

@inproceedings {mohoney_quake_2025,
author = {Jason Mohoney and Devesh Sarda and Mengze Tang and Shihabur Rahman Chowdhury and Anil Pacaci and Ihab F. Ilyas and Theodoros Rekatsinas and Shivaram Venkataraman},
title = {Quake: Adaptive Indexing for Vector Search},
booktitle = {19th USENIX Symposium on Operating Systems Design and Implementation (OSDI 25)},
year = {2025},
isbn = {978-1-939133-47-2},
address = {Boston, MA},
pages = {153--169},
url = {https://www.usenix.org/conference/osdi25/presentation/mohoney},
publisher = {USENIX Association},
month = jul
}

@inproceedings{
simhadri2026results,
title={Results of the Big {ANN}: Neur{IPS}{\textquoteright}23 competition},
author={Harsha Vardhan simhadri and Martin Aum{\"u}ller and Matthijs Douze and Dmitry Baranchuk and Amir Ingber and Edo Liberty and George Williams and Ben Landrum and Magdalen Dobson Manohar and Mazin Karjikar and Laxman Dhulipala and Meng Chen and Yue Chen and Rui Ma and Kai Zhang and Yuzheng Cai and Jiayang Shi and Weiguo Zheng and Yizhuo Chen and Jie Yin and Ben Huang},
booktitle={The Thirty-ninth Annual Conference on Neural Information Processing Systems Datasets and Benchmarks Track},
year={2026},
url={https://openreview.net/forum?id=dB6W56wQL9}
}

@article{xu2025place,
  title={In-place updates of a graph index for streaming approximate nearest neighbor search},
  author={Xu, Haike and Manohar, Magdalen Dobson and Bernstein, Philip A and Chandramouli, Badrish and Wen, Richard and Simhadri, Harsha Vardhan},
  journal={arXiv preprint arXiv:2502.13826},
  year={2025}
}

@inproceedings{wang2021fast,
  title={Fast k-nn graph construction by gpu based nn-descent},
  author={Wang, Hui and Zhao, Wan-Lei and Zeng, Xiangxiang and Yang, Jianye},
  booktitle={Proceedings of the 30th ACM International Conference on Information \& Knowledge Management},
  pages={1929--1938},
  year={2021}
}

@article{liu2026gpu,
  title={GPU-Accelerated Algorithms for Graph Vector Search: Taxonomy, Empirical Study, and Research Directions},
  author={Liu, Yaowen and Chen, Xuejia and Tian, Anxin and Li, Haoyang and Li, Qinbin and Zhang, Xin and Zhou, Alexander and Zhang, Chen Jason and Li, Qing and Chen, Lei},
  journal={arXiv preprint arXiv:2602.16719},
  year={2026}
}

@InProceedings{Guo2026,
  author    = {Guo, Hao and Lu, Youyou},
  booktitle = {24th USENIX Conference on File and Storage Technologies (FAST 26), Santa Clara, CA},
  title     = {Odinann: Direct insert for consistently stable performance in billion-scale graphbased vector search},
  year      = {2026},

}

@inproceedings{xu_spfresh_2023,
  author    = {Yuming Xu and Hengyu Liang and Jin Li and Shuotao Xu and Qi Chen and Qianxi Zhang and Cheng Li and Ziyue Yang and Fan Yang and Yuqing Yang and Peng Cheng and Mao Yang},
  title     = {{SPFresh}: Incremental In-Place Update for Billion-Scale Vector Search},
  booktitle = {Proceedings of the 29th Symposium on Operating Systems Principles (SOSP '23)},
  pages     = {545--561},
  year      = {2023},
  publisher = {ACM},
  doi       = {10.1145/3600006.3613166}
}

@inproceedings{zhao_sivf_2026,
  author    = {Dongfang Zhao},
  title     = {{SIVF}: {GPU}-Resident {IVF} Index for Streaming Vector Analytics},
  booktitle = {The 35th International Symposium on High-Performance Parallel and Distributed Computing (HPDC '26)},
  location  = {Cleveland, OH, USA},
  month     = jul,
  year      = {2026},
  pages     = {14},
  publisher = {ACM},
  address   = {New York, NY, USA},
  doi       = {10.1145/3806645.3807575}
}

@inproceedings{indyk1998ann,
  author    = {Piotr Indyk and Rajeev Motwani},
  title     = {Approximate Nearest Neighbors: Towards Removing the
               Curse of Dimensionality},
  booktitle = {Proceedings of the Thirtieth Annual ACM Symposium on
               Theory of Computing},
  series    = {STOC '98},
  pages     = {604--613},
  year      = {1998},
  doi       = {10.1145/276698.276876}
}

@article{wang2021graphsurvey,
  author  = {Mengzhao Wang and Xiaoliang Xu and Qiang Yue and Yuxiang Wang},
  title   = {A Comprehensive Survey and Experimental Comparison of
             Graph-Based Approximate Nearest Neighbor Search},
  journal = {Proceedings of the VLDB Endowment},
  volume  = {14},
  number  = {11},
  pages   = {1964--1978},
  year    = {2021},
  doi     = {10.14778/3476249.3476255}
}

@inproceedings{milvus2021,
  author    = {Jianguo Wang and Xiaomeng Yi and Rentong Guo and Hai Jin
               and Peng Xu and Shengjun Li and Xiangyu Wang and
               Xiangzhou Guo and Chengming Li and Xiaohai Xu and others},
  title     = {Milvus: A Purpose-Built Vector Data Management System},
  booktitle = {Proceedings of the 2021 International Conference on
               Management of Data},
  series    = {SIGMOD '21},
  pages     = {2614--2627},
  year      = {2021},
  doi       = {10.1145/3448016.3457550}
}

@inproceedings{lewis2020rag,
  author    = {Patrick Lewis and Ethan Perez and Aleksandra Piktus and
               Fabio Petroni and Vladimir Karpukhin and Naman Goyal and
               Heinrich K{\"u}ttler and Mike Lewis and Wen{-}tau Yih and
               Tim Rockt{\"a}schel and Sebastian Riedel and Douwe Kiela},
  title     = {Retrieval-Augmented Generation for Knowledge-Intensive
               NLP Tasks},
  booktitle = {Advances in Neural Information Processing Systems},
  volume    = {33},
  pages     = {9459--9474},
  year      = {2020}
}

@inproceedings{covington2016youtube,
  author    = {Paul Covington and Jay Adams and Emre Sargin},
  title     = {Deep Neural Networks for YouTube Recommendations},
  booktitle = {Proceedings of the 10th ACM Conference on Recommender
               Systems},
  series    = {RecSys '16},
  pages     = {191--198},
  year      = {2016},
  doi       = {10.1145/2959100.2959190}
}

@Article{Jiang2025,
  author    = {Jiang, Haodi and Guo, Hao and Xie, Minhui and Shu, Jiwu and Lu, Youyou},
  journal   = {Proceedings of the ACM on Management of Data},
  title     = {High-Throughput, Cost-Effective Billion-Scale Vector Search with a Single GPU},
  year      = {2025},
  number    = {6},
  pages     = {1--27},
  volume    = {3},
  publisher = {ACM New York, NY, USA},
}

@InProceedings{Gui2026,
  author    = {Gui, Yuntao and Yin, Peiqi and Yan, Xiao and Zhang, Chaorui and Zhang, Weixi and Cheng, James},
  booktitle = {Proceedings of the 32nd ACM SIGKDD Conference on Knowledge Discovery and Data Mining V. 1},
  title     = {PilotANN: Memory-bounded GPU acceleration for vector search},
  year      = {2026},
  pages     = {348--358},
}

@inproceedings{10.1145/3627535.3638475,
author = {Manohar, Magdalen Dobson and Shen, Zheqi and Blelloch, Guy and Dhulipala, Laxman and Gu, Yan and Simhadri, Harsha Vardhan and Sun, Yihan},
title = {ParlayANN: Scalable and Deterministic Parallel Graph-Based Approximate Nearest Neighbor Search Algorithms},
year = {2024},
isbn = {9798400704352},
publisher = {Association for Computing Machinery},
address = {New York, NY, USA},
url = {https://doi.org/10.1145/3627535.3638475},
doi = {10.1145/3627535.3638475},
booktitle = {Proceedings of the 29th ACM SIGPLAN Annual Symposium on Principles and Practice of Parallel Programming},
pages = {270–285},
numpages = {16},
location = {Edinburgh, United Kingdom},
series = {PPoPP '24}
}

@article{lu2025multimodal,
  title={Multimodal data storage and retrieval for embodied ai: A survey},
  author={Lu, Yihao and Tang, Hao},
  journal={arXiv preprint arXiv:2508.13901},
  year={2025}
}

@article{stonebraker2024goes,
  title={What Goes Around Comes Around... And Around...},
  author={Stonebraker, Michael and Pavlo, Andrew},
  journal={ACM Sigmod Record},
  volume={53},
  number={2},
  pages={21--37},
  year={2024},
  publisher={ACM New York, NY, USA}
}


\end{document}